%% file: AManuscript_highlighted_changes.tex
\documentclass[10pt,a4paper]{article}
\usepackage{geometry}
\usepackage{amsmath,amssymb}

\usepackage{changepage}
\usepackage{lscape} 
\usepackage{footnote}
\makesavenoteenv{tabular}
\makesavenoteenv{table}
\usepackage[utf8x]{inputenc}
\usepackage{import}

\usepackage{textcomp,marvosym}

\usepackage{cite}
\usepackage{xr} 
\usepackage{nameref,hyperref}

\usepackage[right]{lineno}

\usepackage{microtype}
\usepackage{epstopdf} 
\DisableLigatures[f]{encoding = *, family = * }

\usepackage[table]{xcolor}
\usepackage{capt-of}
\usepackage{arydshln}
\usepackage{array}
\newcolumntype{P}[1]{>{\centering\arraybackslash}p{#1}}

\usepackage[most]{tcolorbox}
\usepackage{mciteplus}
\usepackage{ragged2e}
\usepackage[font={small,it}]{caption}
\newcolumntype{+}{!{\vrule width 2pt}}

\newlength\savedwidth

\usepackage[aboveskip=1pt,labelfont=bf,labelsep=period,justification=raggedright,singlelinecheck=off]{caption}

\makeatletter
\renewcommand{\@biblabel}[1]{\quad#1.}
\makeatother

\usepackage{lastpage,fancyhdr,graphicx}
\usepackage{epstopdf}
\usepackage{tabularray}
\usepackage{arydshln}
\fancyheadoffset[L]{2.25in}
\fancyfootoffset[L]{2.25in}
\usepackage{color}

\definecolor{dbluecolor}{rgb}{0,0,0}
\def\dBlue#1{{\color{dbluecolor} #1}}

\makeatletter
\newcommand*{\addFileDependency}[1]{
  \typeout{(#1)}
  \@addtofilelist{#1}
  \IfFileExists{#1}{}{\typeout{No file #1.}}
}
\makeatother

\usepackage{floatrow}
\DeclareFloatFont{tiny}{\tiny}
\begin{document}
\nolinenumbers
\vspace*{0.2in}

\begin{flushleft}
{\Large
\textbf\newline{Modelling \dBlue{daily} weight variation in honey bee hives} 
}
\newline
\\
Karina Arias-Calluari\textsuperscript{1,2} \Yinyang,
Theotime Colin\textsuperscript{2}\Yinyang,
Tanya Latty\textsuperscript{2},
Mary Myerscough\textsuperscript{1},
Eduardo G. Altmann\textsuperscript{1}
\\
\bigskip
\textbf{1} School of Mathematics and Statistics, The University of Sydney, Sydney, New South Wales, Australia
\\
\textbf{2} School of Life and Environmental Sciences, The University of Sydney, Sydney, New South Wales, Australia
\\
\bigskip

%
%
\Yinyang These authors contributed equally to this work.


\end{flushleft}
\section*{Abstract}

\justifying
A quantitative understanding of the dynamics of bee colonies is important to support global efforts to improve bee health and enhance pollination services. Traditional approaches focus either on theoretical models or data-centred statistical analyses. 
Here we argue that the combination of these two approaches is essential to obtain interpretable information on the state of bee colonies and show how this can be achieved in the case of time series of intra-day weight variation.  
We model how the foraging and food processing activities of bees affect global hive weight through a set of ordinary differential equations and show how to estimate \dBlue{reliable ranges for} the ten parameters of this model from measurements \dBlue{ on a single day}. Our analysis of 10 hives at different times shows that crucial indicators of the health of honey bee colonies are estimated robustly and fall in ranges compatible with previously reported results. The indicators include the amount of food collected (foraging success) and the number of active foragers, which may be used to develop early warning indicators of colony failure.
\section*{Author summary}
\begin{tcolorbox} [colback=gray!10]
Honey bees are under threat and dying at an alarming rate due to pesticides, parasites, and other stressors. Obtaining information about the health of bee colonies is essential to understand how this happens and to identify measures that can prevent this from happening. Herein, we built a mathematical model of \dBlue{the daily dynamics of hives} that allows such information to be extracted without detailed and expensive measurements. Based only on measurements of how the weight of the hive changes during the day, our model can be used to estimate how many bees are collecting food, how successful they are, and how much time they spend outside the hive. Due to its simplicity, the model presented here can be applied to a wide range of hive scale systems and help beekeepers track how healthy and productive their bees are.
\end{tcolorbox}


\section{Introduction}
\label{}

There is growing concern that crop pollination by honey bees might no longer be a sustainable operation in the near future \cite{goulson2015combined}. Single and combined stress from pesticides, parasites and diseases have led to an increase in bee colony mortality worldwide\cite{goulson2015bee, hladik2018environmental,siviter2021agrochemicals,le2010varroa,powney2019widespread,woodcock2017country}. Despite beekeeper’s efforts to control pests and diseases, annual colony losses of about 30\% are frequently reported in countries across Europe \cite{corbet1991bees} and North America\cite{lee2015national,currie2010honey,jacques2017pan}. 
Automated methods to track the status of hives can play a key role in helping beekeepers prevent the death of colonies\cite{meikle2015application,francis2013varroa,morawetz2019health,horn2021honey,charreton2015locomotor}.

The social organization of a bee colony is complex and relies on positive and negative feedback loops to maintain homeostasis. Feedback loops can help buffer the effect of stressors or cause snowballing effects which in some cases precipitate colony death. A well-known feedback loop that has been hypothesized to accelerate the failure of the colony is the one that regulates the ontogeny of foraging\cite{leoncini2004worker}. In a bee colony, workers typically spend the first few days of their adult life performing various maintenance, defence, and care tasks before transitioning into foragers\cite{seeley1991age}. The presence of sufficiently large number of foragers inhibits the transition of younger bees into foragers\cite{robinson1987regulation, leoncini2004worker}. Under ideal conditions, this ensures enough bees are performing in-hive tasks, and that the colony is able to process all the food brought back by the foragers and to raise the next generation of workers. If an environmental stress reduces the number of foragers, this same mechanism ensures the rapid replacement of the foraging force. With fewer foragers, the transition of younger bees into foragers is increased and the foragers' population is rapidly replenished. 
Under sustained stress however, it has been hypothesized that this buffering mechanism can lead to generations of workers starting to forage at a younger age, which reduces their life expectancy and the foragers' population increasingly fast until colony death\cite{perry2015rapid}. Thus, information about the number of forager bees and their activities can provide important guidance about the health of the hive. 

\begin{figure}[!ht]
\centering
\includegraphics[scale=0.7,trim=4cm 7cm 4cm 7cm, angle =0 ]{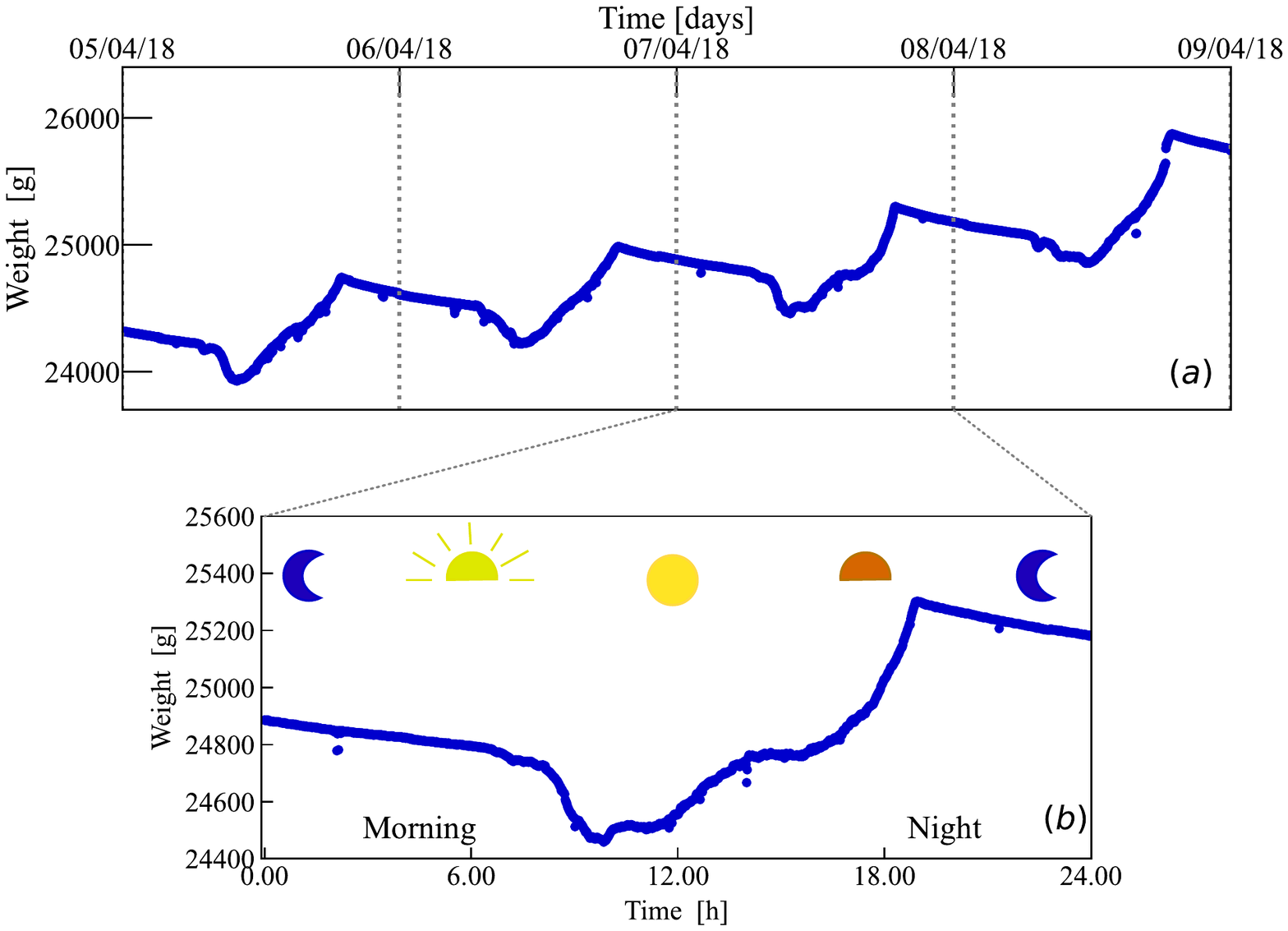}
\caption{(a) The weight of a bee hive shows systematic variations throughout a day. The blue dots show the weight of a hive measured every minute for four consecutive days. (b) Each daily hive weight's curve present consistent patterns at specific time ranges of morning, afternoon, evening and night time.} 
\label{fig:1}
\end{figure}

Directly measuring the number of forager bees is difficult, but the use of new technologies provides opportunities to estimate them from more accessible data \cite{henry2012common,meikle2015application}. Recently developed affordable and precise balances can easily be fitted to hives in the field to study the long-term weight changes\cite{quinlan2022grassy} and shorter within-day weight variations of hives\cite{meikle2015application, meikle2016monitoring}.
Within-day hive weight variations are clearly visible and exhibit similar patterns that are consistent across multiple day as shown in Fig.~\ref{fig:1}. In the morning, rapid weight loss occurs because foragers leave the hive \cite{meikle2018using}. For this reason, previous methods have focused on measuring the slope and low point of the morning departure segment of daily hive weight’s curves\cite{meikle2018using}, and have used this measure as likely representing forager activity\cite{colin2019long}. However, these phenomenological methods provide limited information about the daily internal dynamics of the hive. Other methods perform piece-wise linear regression analysis where up to five breaking points are suggested after an abrupt change of slope \cite{meikle2018using,meikle2008within}. In particular, these methods do not provide quantitative estimations of the relevant bee activities such as the number of forager bees or the amount of food brought to the hive because the different bee activities are not explicitly modeled and because the simultaneous departures and returns of bees prevent trivial calculation of the population of forager bees.

In this paper, we develop a mechanistic mathematical model \dBlue{of the daily dynamics of hives} which, in combination with hive weight time series, \dBlue{provides} quantitative estimations of key indicators of the state of the hive such as the total number of active forager bees and the average mass of food they bring to the hive in each trip.
We start in Sec.~\ref{sec.model} with our model presentation, its qualitative description, and its main properties. In Sec.~\ref{sec.inference} we present how the parameters can be inferred from data. In Sec.~\ref{sec.results} we apply our methodology to time series of ten different hives measured across a year and show how in most cases the results lead to biologically meaningful estimations. Finally, in Sec.~\ref{sec.conclusions} we summarize our conclusions. A detailed description of our data and computational methods is presented in the Materials and Method section. The data and codes used in our analysis are available in a repository~\cite{GitHub}.

\section{Results}

\subsection{Model}\label{sec.model}

Our aim is to obtain a mechanistic model that contains the essential processes necessary to describe how the weight $W$ of a bee hive changes throughout time $t$ over a daily cycle. %

\begin{figure}[h]
\centering
\includegraphics[scale=0.5,trim=1cm 1cm 1cm 1cm, angle =0 ]{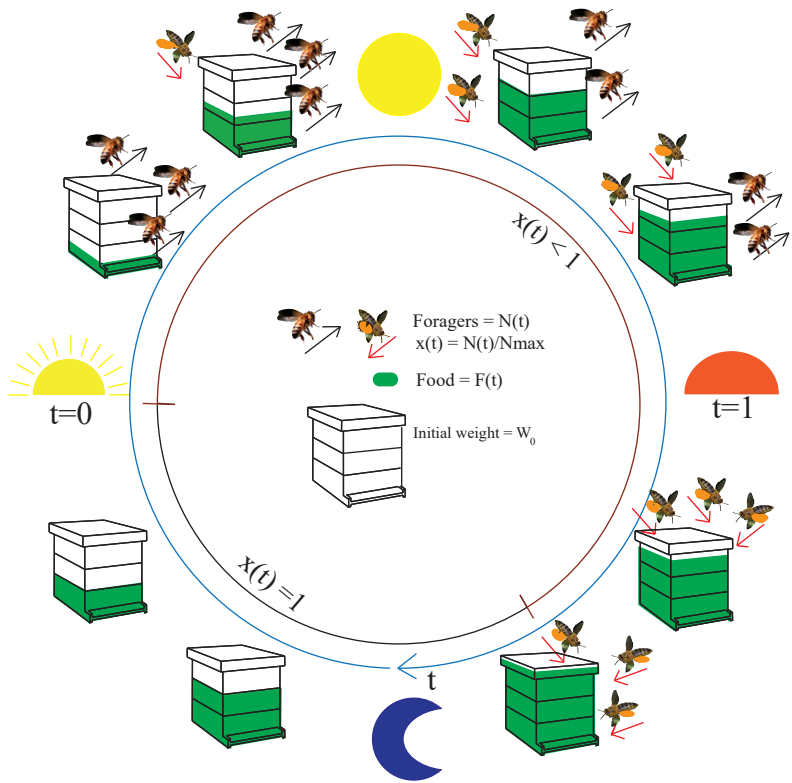}
\caption{\dBlue{Schematic representation of the daily cycle of bee activities that impact in modeling colony weight, where time flows in a clockwise direction (blue line). The main time-dependent elements we model are the number of forager bees $N(t)$ and the food $F(t)$ (in green) inside the hive. The foragers are represented as active bees that depart the hive (black arrow) and return to it (red arrow) with food (in orange). The ratio of forager bees inside the hive at a particular time is represented by $x(t)$, which is constant before and after the dynamics of forager bees because the $N_{max}$ foragers are all inside the hive. The static weight $W_0$ is composed of the hive structure and bees that do not leave the hive. }}
\label{fig:2}
\end{figure}

\paragraph{Qualitative description.} Our model will incorporate the following well-known processes and honey-bee behaviours that are also schematically illustrated in  \dBlue{Fig~\ref{fig:2}}:

\begin{enumerate}
    \item In the morning, the weight of the hive rapidly falls as foragers leave the hive\cite{meikle2018using}. This initial weight drop usually lasts for a few hours, indicating that the weight loss is only partly counterbalanced by forager bees returning to the hive. A successful forager will return with pollen, nectar, water or propolis. The weight of a forager bee is expected to be higher when they return than when they leave, if the resources they seek are available in their environment. 

    \item Around midday, departures and arrivals reach roughly an equilibrium. Hives in environments with food resources start gaining weight, possibly indicating that the weight of departing foragers is less than the weight of effective foragers returning with food to the colony\cite{meikle2018using}. The weight increases until all foragers have returned at dusk.

    \item In the late afternoon, the number of departures reduce until they stop completely around dusk. Once all the foragers have returned, the hive \dBlue{stops} gaining weight.
    
    \item At night, the weight of a hive declines slowly untill dawn due to \dBlue{respiration, food consumption, and evaporation}. The metabolic activity of a bee colony remains high at night because bees actively maintain the temperature of the brood at about 35 degrees, as well as humidity and $CO_2$ concentration to optimize development. Bees actively warm the colony by shivering, or cool down and ventilate the hive by evapotranspiration and fanning. These activities require a great deal of movement and, thus, causes honey bees to consume significant amounts of honey at night. Sugars in the honey are broken down to produce energy and exhaled as $CO_2$. A colony also  evaporates a large amount of water, either to facilitate biological processes, concentrate sugars in food stores, or cool down the hive. Moreover, it is likely that a small amount of the weight loss at night results from cleaning activities, with bees likely getting rid of pests, dead bees and chewed brood cell caps throughout the night. The cycle resumes the next day again.
\end{enumerate}

The net outcome of the activities described above depends on the hive, the environment, and the \dBlue{weather conditions}. In favourable environments and at its peak population in summer, a single bee colony can harvest a few kilograms of food in a single day. Most of this weight comes in the form of nectar and honeydew, which generally contains more than 50\% water and diluted sugars\cite{chalcoff2006nectar}. Hives in environments with little or no resources may \dBlue{lose} more weight than they gain in a day as a consequence of honey consumption inducing respiration and evaporation (i.e., foragers expend more resources than they are able to bring back to the colony\dBlue{)}. 

\paragraph{Mathematical model.} We consider the time dependent weight $W \in \mathbb{R}^+$ \dBlue{(measured in $g$)} to depend on the number of forager bees $N \in \mathbb{Z}, 0 \le N \le N_{max}$ \dBlue{(measured in bees)} and on the food $F \in \mathbb{R}^+$ \dBlue{(measured in g)} as
\begin{equation}\label{eq.W}
W(t)=F(t)+wN(t)+W_0,
\end{equation}
where $W_0$ is the constant weight of the hive and $w$ is the (average) weight of a bee. Next we discuss the variation in $N(t)$ and $F(t)$. 

For mathematical convenience, we model the variation in number of foragers bees $N(t)$ by focusing on the fraction $x \in \mathbb{R}$ (nondimensional) of all $N_{max}$ foragers bees  that are inside the hive:
\begin{equation} \label{eq.xdef}
  x\equiv \frac{N}{N_{max}} \Rightarrow N(t) = N_{max} x(t).
\end{equation}
We write the rate of change of $x$ as
\begin{equation}\label{eq.x}
\frac{dx}{dt} = a(t) (1-x) - d(t) x,
\end{equation}
where $a(t)$\dBlue{, measured in $1/s$,} is the {\it a}rrival rate of the fraction of forager bees outside the hive and $d(t)$ \dBlue{, measured in $1/s$,} is the {\it d}eparture rate of the fraction of forager bees inside the hive. The inverse of these rates can be thought of as the typical period a bee spends outside and inside the hive, respectively.  These terms multiply the fraction of bees outside ($1-x$) and inside ($x$), respectively. The rate of bees arriving is
$$A(t)=a(t)(N_{max}-N(t))= a(t) N_{max}(1-x(t))$$
and the rate of bees departing is
$$D(t)=d(t)N(t)= d(t) N_{max}x(t),$$
where $A$ and $D$ are \dBlue{measured in $bees/s$}.

\dBlue{The food inside the hive $F(t)$ increases due to the mass $m$ (measured in $g$) of food brought in by arriving bees and decreases due to evaporation, respiration, and food consumption, which we model as a constant loss $\ell$ (measured in $g/s$)}\footnote{\dBlue{More sophisticated models could consider a time dependent loss (e.g., due to temperature or weather variations) and a term proportionally to $N$ (or $x$), indicating a rate of consumption proportional to the number of (forager) bee. }} \dBlue{so that}

\begin{equation}\label{eq.F}
    \frac{dF}{dt} = - \ell + m A(t) = N_{max}( m a(t)(1-x))-\ell.
\end{equation}

Notice that $F(t)$ is linear in $x$ and can be computed directly from $x(t)$ and $a(t)$ by integrating the above equation. Substituting $F(t)$ and $x(t)$ in Eq.~(\ref{eq.W}) we obtain the time-dependent weight $W(t)$ of the hive, our measurable quantity. This implies that the equations can be solved sequentially, starting from $x(t)$. The final step of our model is to specify the departure $d(t)$ and arrival $a(t)$ rates defined in Eq.~(\ref{eq.x}) and then solve for $x(t)$. We will consider a very simple choice for $a(t)$ and $d(t)$ -- piecewise constant in $t$ -- which allows for an explicit solutions for $x(t)$ and, thus, $W(t)$. More sophisticated models could use more elaborated functions $a(t)$ and $d(t)$ or even delay equations. 

\paragraph{Day and night cycles.} Without loss of generality, in our model, we consider:
\begin{itemize}
\item[$t=0$] to be the time in the morning (around sunrise) at which the bees start exiting the hives.
\item[$t=1$] to be the time in the late afternoon (around sunset) when bees stop leaving the hive.
\end{itemize}
Therefore, in all cases below we assume $d(t)=0$ for $t<0$ and $t>1$. The matching of the model time $t$ to the clock time requires the estimation from data of two clock times: $t_0$ (corresponding to $t=0$) and $t_1$ (corresponding to $t=1$).

\paragraph{Night dynamics.} Let us start with the (easiest) night case $t> 1$, when bees do not depart $d(t)=0$ and arrive at a constant rate $a(t)=a$. Assuming that at the beginning of this regime at $t=1$ there are $x_1$ bees inside, the particular solution of Eq.~(\ref{eq.x}) is
\begin{equation}\label{eq.xnight}
x(t) = 1-(1-x_1)e^{-a(t-1)},
\end{equation}
which corresponds to an exponential decay towards all the bees being inside the hive ($x=1$) with a rate $a$. Therefore, we can consider $1/a$ proportional to the duration for which bees remain outside the hive.

\paragraph{Day dynamics.} Let us consider the simple case where $d(t)=d$ and $a(t)=a$, both constant in time. The initial condition in this case is $x=1$ at $t=0$ because we assume all forager bees spent the night in the hive (i.e., $1/a \ll$ night period and all previous-day foragers return). In this case, Eq.~(\ref{eq.x}) leads to
\begin{equation}\label{eq.xday}
x(t) = \frac{a}{d+a} + \frac{d}{d+a}e^{-(d+a)t}.
\end{equation}
\paragraph{Explicit solutions.} We are now able to combine the results above to write explicit solutions for the fraction of foragers inside the hive $x(t)$, the weight of food inside the hive $F(t),$ and the total weight of the hive $W(t)$. 
We consider the arrival rates during the day $a_1$ and at dusk $a_2$ to be different:
\begin{equation}\label{eq.as}
a(t) = \left\{\begin{array}{ll} 
a_{1}& \text{ for }  t \le 1,\\
a_{2} & \text{ for } t>1,
\end{array}\right.
\end{equation}
where we expect to retrieve $a_2>a_1$ (bees spend less time out when it is getting dark).

Introducing these arrival rates and a constant departure rate $d$ in Eqs.~(\ref{eq.xnight}) and (\ref{eq.xday}) we find an explicit solution divided in three time intervals for $x(t)$:
\begin{equation}\label{eq.x2}
x(t) = \left\{\begin{array}{ll} 
1 & \text{ for } t<0, \\
\frac{a_1}{d+a_1} + \frac{d}{d+a_1}e^{-(d+a_1)t} & \text{ for }  0 \le t \le 1,\\
1-(1-x_1)e^{-a_2(t-1)} & \text{ for } t>1
\end{array}\right.
\end{equation}
where $x_1=\frac{a_1}{d+a_1} + \frac{d}{d+a_1}e^{-(d+a_1)}$ is the value of $x(t=1)$ and ensures the continuity of $x(t)$. 

We can now turn to the dynamics of food inside the hive $F(t)$. From Eq.~(\ref{eq.F}) we obtain
$$ F(t) = C + (\tilde{m} -\ell)t- \tilde{m} \int x(t) dt, $$
where $\tilde{m}=mN_{max}a$  and $C$ is an integration constant to be fixed using the initial condition $F(0)=F_0$ and $F(1)=F_1$. The solution for $F(t)$ is then solved by integrating Eq.~(\ref{eq.x2}), which can be done analytically in each of the three branches leading us to
\begin{equation}\label{eq.F2}
F(t) = \left\{\begin{array}{ll} 
F_0 - \ell t & \text{ for } t<0, \\
F_0 - \frac{\tilde{m} d}{(d+a_1)^2}+(\tilde{m} (1-\frac{a_1}{d+a_1})-\ell)t + \tilde{m} \frac{d}{(d+a_1)^2}e^{-(d+a_1)t} & \text{ for }  0 \le t \le 1,\\
F_1 + \ell +\tilde{m}\frac{1-\textcolor{black}{x_1}}{a_2}-\ell t-\tilde{m} \frac{1-\textcolor{black}{x_1}}{a_2} e^{-a_2(t-1)} & \text{ for } t>1,
\end{array}\right.
\end{equation}
where $F_0\equiv F(t=0)$ -- for this model an arbitrary constant that is incorporated into the static weight of the hive $W_0$ in Eq.~(\ref{eq.W}), consequently, without loss of generality, we take $F_{0} =0$ -- and $F_1\equiv F(t=1)$ is computed setting $t=1$ in the second line of Eq.~(\ref{eq.F2}) so that 
$$ F_1= F_0 - \frac{\tilde{m}d}{(d+a_1)^2}+\left(\tilde{m} \left(1-\frac{a_1}{a_1+d}\right)-\ell\right) + \frac{\tilde{m} d}{(d+a_1)^2}e^{-(d+a_1)}.$$
Finally, the weight of the hive $W(t)$ is obtained from Eq.~(\ref{eq.W}) as the sum of Eq.~(\ref{eq.x2}) (times $w N_{max}$), Eq.~(\ref{eq.F2}), and the constant weight of the hive $W_{0}$, leading to an explicit piece-wise continuous function $W(t)$ that will be compared to the measured data. A summary of the parameters used in the model is given in Table~\ref{tab.parameters}.

\begin{table}[!ht]
\small
\begin{tabular}{lll}
\hline
\textbf{Parameter} & \textbf{Description} & \textbf{Prior} \\
 &  & \textbf{Range} \\
 \hline
$w:$& the average weight of a single bee, introduced in Eq.~(\ref{eq.W}).  & 0.113 g/bee \cite{thompson2016extrapolation,thompson2019honeybees}\\
$W_0:$ & the static weight of the hive that includes the & $[5000  \,,\, \dBlue{max(W)}]$g\\
& structure, food stores, non-foraging bees, etc., see Eq.~(\ref{eq.W})&\\
$N_{max}:$& the total number of foragers active on that day,  & $( 0 \,,\, 80000]$ bee\\
&  introduced in Eq.~(\ref{eq.xdef}) & \\
$\ell:$& \dBlue{the continuous loss}, introduced in Eq.~(\ref{eq.F}) & $( 0 \,,\,\, \dBlue{max(W)}/h]$ g/h\\
$m:$& the average mass of food brought by foragers, introduced in Eq.~(\ref{eq.F}) & $( 0 \,,\,  0.78w]$ g/bee \cite{garcia2004variations, visscher1996honey, afik2006effect}\\
$a_1,a_2:$ & the rates of arrivals, see Eq.~(\ref{eq.as}),& $[ 0.10 \,,\, 4.80]$ 1/h \cite{colin2021effects, he2013assessment, rodney2020dietary}\\
&inversely proportional to the time bees spend out- and &\\
$d:$ & the rates of departures, see Eq.~(\ref{eq.as}),& $[ 0.81 \,,\, 3.01]$ 1/h \cite{colin2021effects, he2013assessment, rodney2020dietary}\\
&inversely proportional to the time bees spend out- and &\\
& in-side the hive, respectively.&\\
$t_0,t_1:$& the clock time corresponding to $t=0$ and $t=1$, respectively& $[6.50 \,\, 9.00]$ h\\
& & $[14.5 \,\, 18.5]$ h\\
\hline
  \end{tabular}
\caption{The parameters $\theta$ of our model. For simplicity, the average weight of bees $w$ is fixed based on previously reported results. The remaining parameters are inferred from data within a prior range of admissible parameters (right column). \dBlue{The boundaries of the range of flat priors were chosen based on the cited references or on values smaller/larger than reasonably possible values (e.g., the maximum weight max(W) is chosen to be the maximum total weight of the hive at all times). In all cases reported below, the inferred parameters remained away from the boundaries, indicating that our choice of boundaries has no influence on our results.}}
\label{tab.parameters}
\end{table}

\paragraph{Effective description of the model solution W(t).} The mathematical model described above is divided in three regimes -- $t<0, 0 \le t \le 1, t>1$ --  with similar characteristics as shown in Fig.~\ref{fig.3}. The three regimes contain segments of linear dependence of $W(t)$ -- a decay in the first and last regime, and typically an increase in the intermediate regime -- with a transition period (shaded region) between the regimes:

\begin{itemize}
    \item[$t<0$] The first regime corresponds to all bees inside the hive $(x=1)$ and a linear decay of $W$ on time $t$ with   rate $\ell$-- evaporation and respiration rate defined in Eq.~(\ref{eq.F})--  and with an intercept $A\equiv W(t=0)$. 
    \item[$0\le t \le 1$] The second regime starts with an exponential relaxation of $x(t)$ from $0$ to the fixed point value $x^* \equiv \frac{a}{d+a}$, as described in Eq.~(\ref{eq.xday}), which leads to a sharp decay of $W(t)$ until an inflection time $t_c$. Shortly after $t_c$, $W(t)$ shows a linear dependency with rate $\alpha$ and constant $B$. 
    \item[$t>1$] The third regime shows an exponential relaxation of $x(t)$ from $x^*$ to $x=1$ as described in Eq.~(\ref{eq.xnight}), which leads to a sharp growth in $W(t)$ as all the bees return and before the linear decay of the first regime starts again completing the cycle. 
\end{itemize}
    
 The critical regime describing bee activities is the second regime that takes place during the day $0\le t \le 1$.  It can be effectively described by the parameters defined in Table \ref{tab.parameters}, which can be rewritten in terms of the robust parameters of the model as:
 
\begin{figure}[!ht]
\centering
\includegraphics[scale=0.33,trim=1cm 5cm 1cm 1cm, angle =0 ]{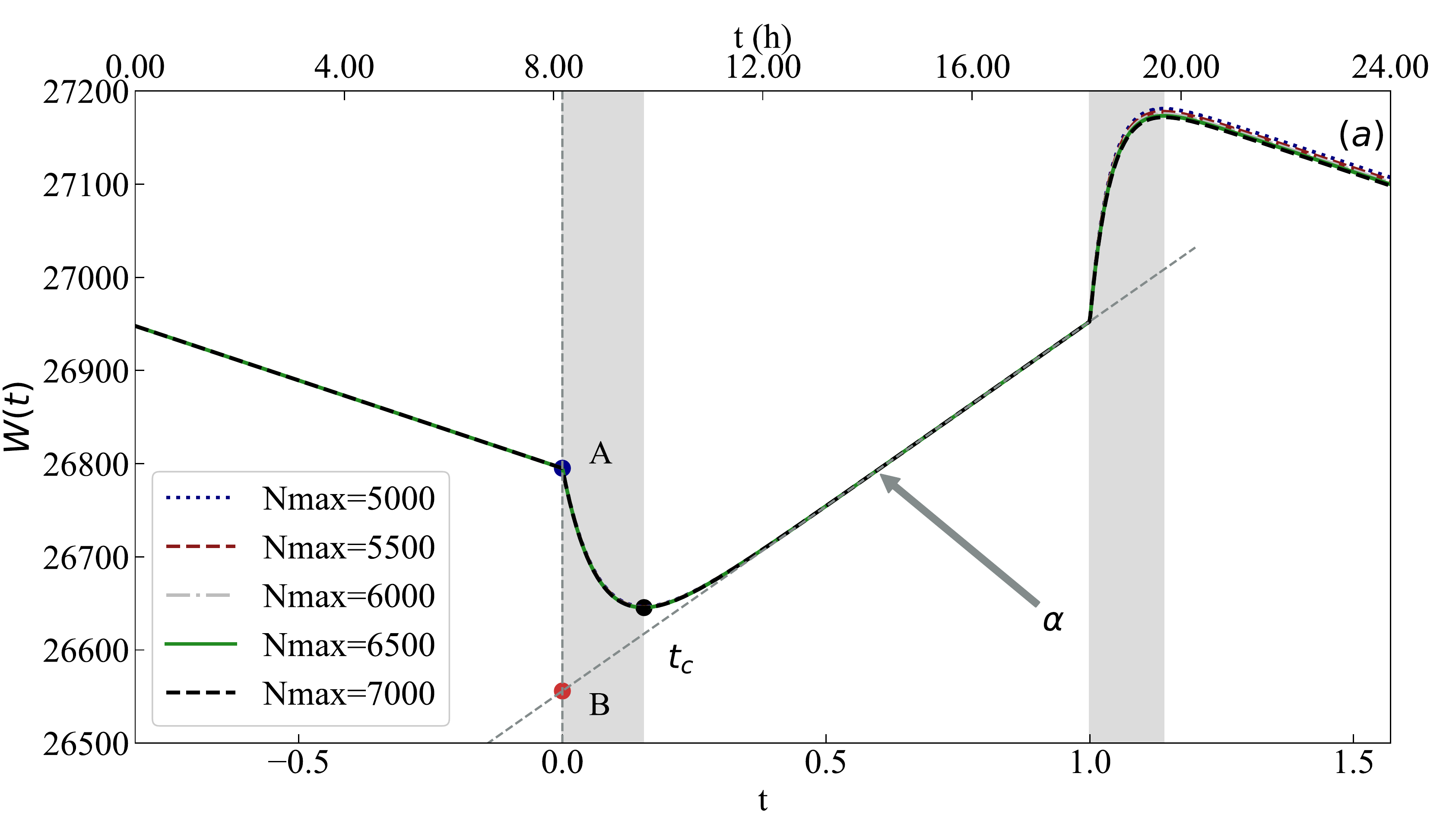}
\includegraphics[scale=0.33,trim=0cm 8cm 1cm 1cm, angle =0 ]{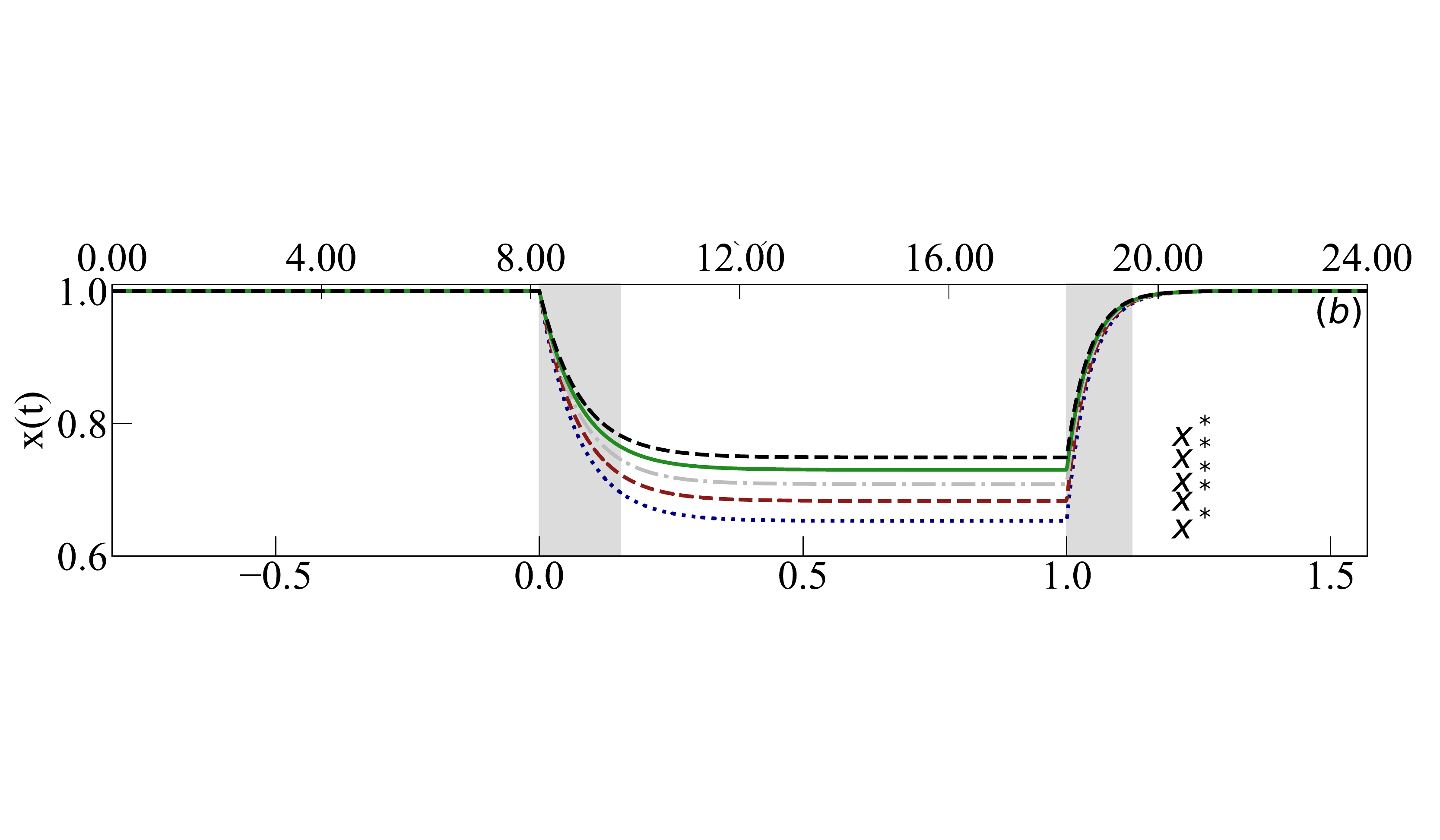}
\includegraphics[scale=0.33,trim=0cm 8cm 1cm 3cm, angle =0 ]{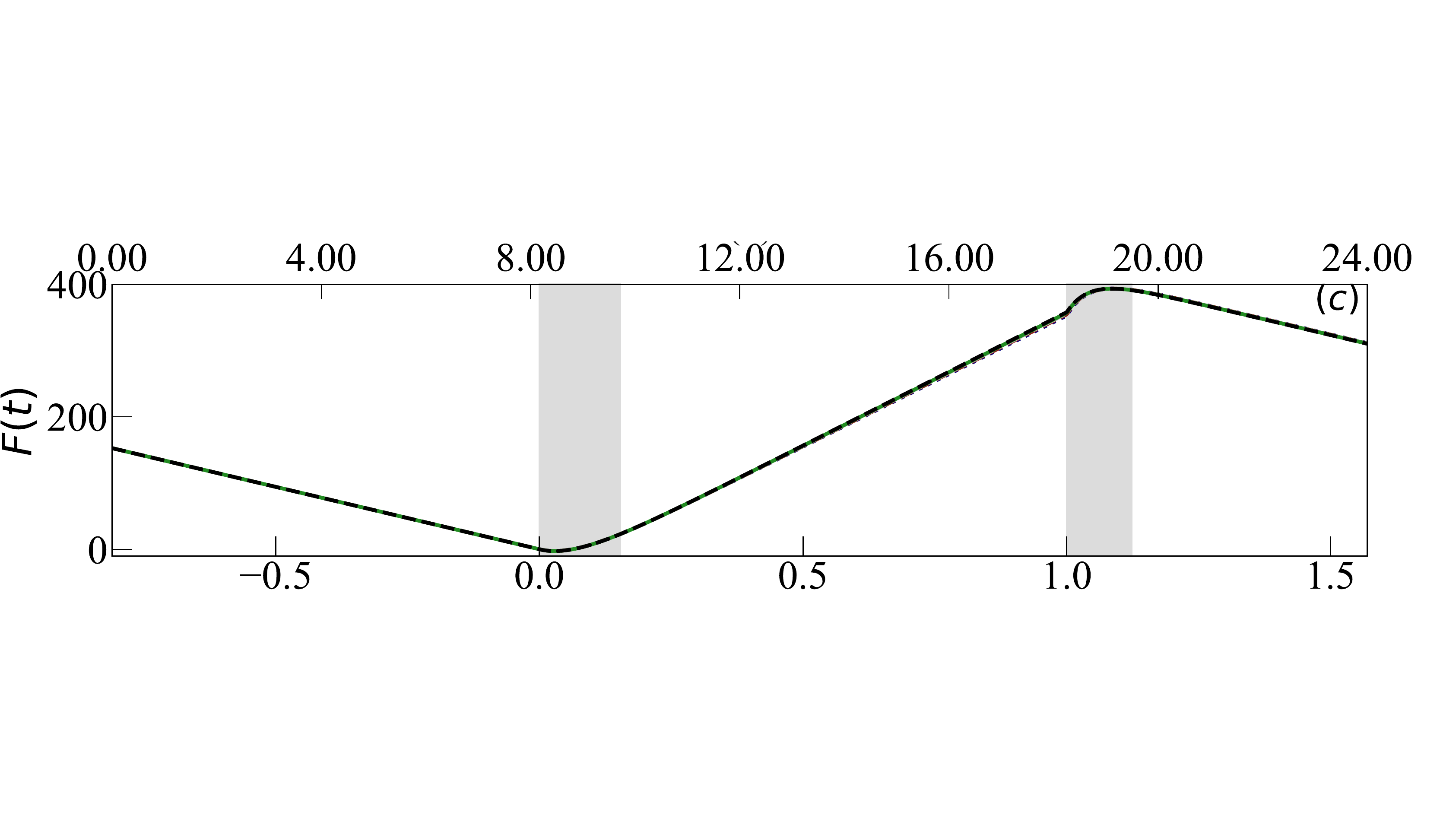}
\includegraphics[scale=0.33,trim=0cm 6cm 1cm 3cm, angle =0 ]{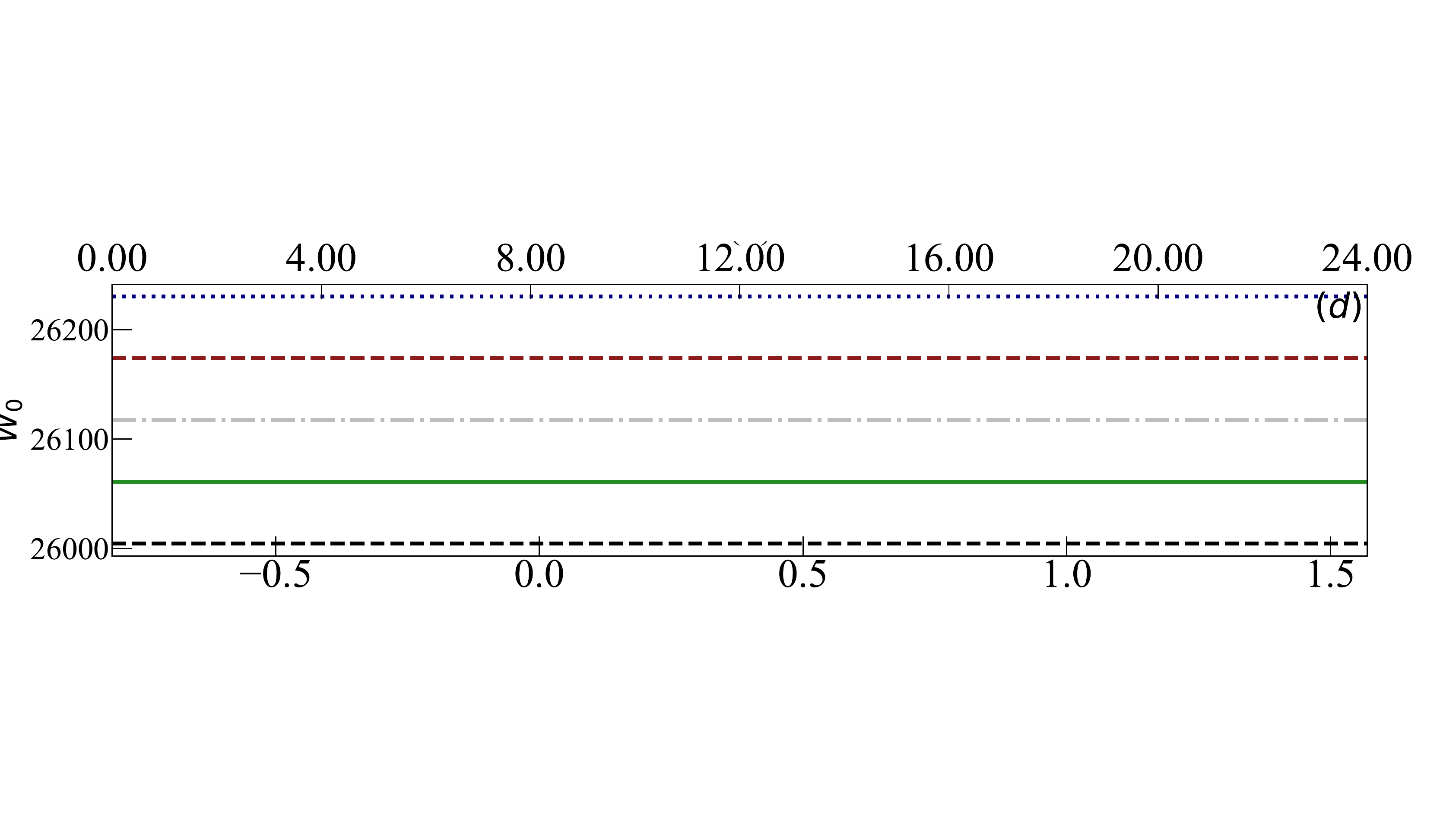}
\caption{Graphical representation of the solution of our model.  (a) The weight as a function of time, $W(t)$, in which the three linear regimes and the two transition regimes (shaded areas) are visible. (b-d) The three key elements that compose $W(t)$: (b) the number of foragers inside the hive $x(t)$ that converges to the constant value $x^*$; (c) the amount of food $F(t)$ inside the hive; and (d) the constant weight of the hive $W_0$. In all four plots, five curves are shown for different model parameters $\theta=\{N_{max},W_0,a_1,a_2,d,m,\ell=18.74 \,\,g/h, w=0.113\,\,g/bee \}$ that lead to the same effective parameters $A,B,t_c, \alpha$ shown in panel (a). The values of the remaining parameters in numerical order for $N_{max}$ are $W_0[g]=26230,\,\, 26173,\,\, 26117,\,\, 26060,\,\, 26004;\,\,  a_1[1/h]=0.882,\,\,0.914,\,\,0.941,\,\,0.964,\,\,,0.984;\,\, a_2[1/h] =2.329,\,\,2.328,\,\,2.328,\,\,2.327,\,\,2.327;\,\, d[1/h]=0.469,\,\,0.424,\,\,0.387,\,\,0.356,\,\,0.330; and\,\, m[g/bee]=0.038,\,\,0.037,\,\,0.036,\,\,0.035,\,\,0.034$. \dBlue{The two grey spans highlight the time intervals of exponential decay and growth in $x(t)$, respectively.}}  \label{fig.3}
\end{figure}
\begin{itemize}
\item[A:] The total weight of the hive at $W({t=0})$ defined as: $A=N_{max}w+W_{0}$
\item[$\alpha$:] The rate of weight gain due to foraging, from Eq.~(\ref{eq.F2}): $\alpha=\tilde{m} ( 1 - \frac{a_{1}}{d+a_{1}} )-\ell$ 
\item[$B:$] The intersection point of weight gain at $t=0$: $B= -(\alpha+\ell) \left(\frac{w}{ma_{1}} + \frac{1}{d+a_1} \right) +A$
\item[$t_c:$] The time of the minimum weight:  $t_c={\frac{-1}{({d + a_1})}\ln \left( \frac{\alpha}{dN_{max}w+\alpha + \ell}\right)}$
\end{itemize}

Fig.~\ref{fig.3} shows an important mathematical property of our model: different choices of parameters lead to nearly identical curves $W(t)$ making our model effectively {\it underdetermined} by $W(t)$. Mathematically, this  {\it underdetermination} happens because the \emph{four} effective parameters introduced above ($A,B, \alpha, t_c$), which provide a simplified quantitative description of the solution $W(t)$, depend on \emph{five} model parameters ($N_{max},W_0,m,a_1,d$, note that $\tilde{m}\equiv m N_{max}$ and the mass of a bee $w$ is taken as fixed). This means that there are different choices of the parameters~$\theta$ of our model that lead to the same four effective parameters describing the second regime $W(t)$ (even if the curves $W(t)$ are mathematically distinct, their difference is indistinguishable in Fig.~\ref{fig.3} and much smaller than the experimental precision of the data). Biologically, this underdetermination appears due to our definition of which bees in the hive count as forager bees. Our model parameters are fixed and can be thought as an average over a spectrum of bees that ranges from very active to those that do only very few forager excursions in a day. Depending on whether we include the least active bees as foragers or not, we obtain different model parameters but the same effective parameters defined above.  For instance, the same initial weight $A\equiv W(t=0)=N_{max} w +W_0$ can be obtained by including  the less active bees as foragers  -- thus increasing $N_{max}$ -- or as part of the non-foragers bees that spend most of the time in the hive -- thus contributing to the constant $W_0$. Similarly, counting the least active bees as foragers reduces the average mass $m$ of food brought to the hive, increases the average time they spend in the hive (i.e., decreasing $d$), decreases the time spent outside the hive (i.e., increase $a_1,a_2$), and leads to a larger fraction of forager bees inside the hive ($x^*$ increases). As shown in Fig.~\ref{fig.3}, solutions of $W(t)$ with very different $N_{max}, W_0$ in panels (b) and (d) -- which imply also different  $d,a_1,a_2, \textrm{and} \,\, m$ -- lead to essentially the same curves for the food $F(t)$ in panel (c) and weight $W(t)$ in panel (a).

\subsection{Connecting model and data}\label{sec.inference}
 Here we show how to combine our model with data to extract information about the state of hives. We describe the theoretical and computational methodology used to infer the model parameters~$\theta$ from time series of hive weights $W'(t)$.
 
\paragraph{Theory.} We assume that the measured weight $W'$ deviates from the actual weight $W$ as
$$W'=W+\varepsilon,$$
where  $\varepsilon$ is an independent Gaussian distributed random variable with zero mean and standard deviation $\sigma$, i.e., $\varepsilon \sim \mathcal{N}(0,\sigma^2)$, that accounts for measurement errors and random effects not accounted in our model.

If $\theta$ are the model parameters generating $W \equiv W_\theta$, at an arbitrary time $t$, as described in the model of the previous section, the probability likelihood of an observation $W'$ is a Gaussian distribution centred at $W_\theta$ given by
$$P(W'|\theta) = \mathcal{N}(W_\theta;\sigma^2).$$
Assuming that observations are performed at times $q_1,q_2,\ldots q_n$ with $q_i=q_0+n \delta q$ and are (conditionally) independent from each other, the log-likelihood of the observations is given by:
\begin{equation}\label{eq.L}
\log P(W'|\theta) = \sum_{i=1}^{n} \log \mathcal{N}(W_\theta(q_i); \sigma) = \sum_{i=1}^{n} -\frac{(W'(q_i)-W_\theta(q_i))^2}{2 \sigma^2}- \log(\sqrt{2 \pi}\sigma)
  \end{equation}
We can include further information on the parameters $\theta$ by considering a prior distribution $P(\theta)$. Here we consider flat priors for all parameters, i.e., $P(\theta) = cst.$ for $\theta \in [\theta_1,\theta_2]$ as specified in Tab.~\ref{tab.parameters} above. Effectively, this restricts the range of admissible parameters without changing the probability of parameters within the range. The estimation of the parameters $\theta$ can be done maximizing the posterior distribution $P(\theta | D)$ which in our case is equivalent to the maximization of the log-likelihood $(\ref{eq.L})$ or to the minimisation of the squares of $W'-W$ as
\begin{equation}
L = \sum_{i=1}^{n} (W'(\tau_i)-W(\tau_i))^2
\end{equation}\label{Eq.Lsq}
in the range of admissible parameters set by the priors.

\paragraph{Computation.} The problem of obtaining the ten parameters introduced in our model -- see Table~\ref{tab.parameters} -- has been thus reduced to the optimization problem of finding the minimun of $L$ in a ten-dimensional space. The computational method we use for this optimization is centred around the Levenberg\textendash Marquardt algorithm as implemented in the library SciPy (See~ Materials and Methods Sec-\ref{App.Computational_methods}). However, a robust estimation of the parameters cannot be obtained directly through the na\"ive application of this method both because of the high-dimensionality of the problem and the effective underdetermination of our model reported at the end of last section (which leads to essentially flat regions of $L$ in the parameter space with multiple local minima). Here we use properties of bee hives and our model to simplify this problem and obtain a robust estimation of parameters.

We start by considering \dBlue{four parameters}:

\begin{itemize}
\item[$w$] For simplicity, we fix the average weight of bees to $w=0.113\,g/bee$\cite{thompson2016extrapolation,thompson2019honeybees}.
\item[$t_0,t_1$] We infer the times $t_0$ and $t_1$ in a grid of plausible values (separated by five minutes). During this minimization we do not distinguish between model parameters that lead to the same effective parameters $A,B,\alpha, \, \, \textrm{and} \,\,t_c$ because $t_0$ and $t_1$ are not involved in them. The estimation of $t_0$ and $t_1$ \dBlue{allow us} to re-scale the data to the model time scale, facilitating the numerical estimation of the remaining parameters.
\item[$\ell$]  \dBlue{The continuous loss} can be calculated directly from the negative linear decrease when $t<0$.
\end{itemize}

These parameters, and the effective parameters ($A,B,\alpha,t_c$), are not strongly affected by the underdetermination problem discussed above and are thus part of our {\it robust} estimation. The remaining six parameters -- $N_{max}, W_0, m, a_1, a_2, d$ -- \dBlue{are directly} connected to the underdetermination problem and need to be estimated with care. We start by obtaining a robust estimation of $A \,\, ,B$,$\,\,\alpha$, and $t_{c}$ by direct minimization of $L$.  The mapping of the effective parameters to the model parameters is done in two different ways:

\begin{itemize}
\item[(i)]  We \dBlue{seek a} range of plausible model parameters determined as the values of $W_0$ and $N_{max}$ that keep the four effective parameters constant and that maintain the parameters $\ell,m,a_1,a_2$ and $d$ \dBlue{within their prior range} (set in Table.~\ref{tab.parameters}). In practice, this is done by defining different linear combinations between $N_{max}$ and $W_{0}$ that are introduced in the   system of six equations that connect the parameters. For each possible $N_{max}$, we obtain a particular $a_1$, $d$, and $m$ by solving the systems of equations given by $A \,\, ,B$ and $t_{c}$. The second rate of arrival, $a_2$, is obtained by the conservation of departed and returned bees. This method leads to the estimation of an {\it interval} of possible parameters.
\item[(ii)]  Alternatively, by pre-defining one parameter, we calculate the $N_{max}$ value directly because the system of equations presents a unique solution. We focus on the departure rate $d$ because \dBlue{of the consistency among different reported studies on the time interval between trips that a forager bee spends inside the hive \cite{colin2022evaluating} and because $d$ can be directly connected to the simplifying modeling decision of dividing bees into foragers and those that do not leave the hive (i.e., we consider as foragers the most active bees so that collectively their average departure rate is $d$, the other bees are considered to effectively stay inside the hive, and their weight is counted as part of $W_0$). In practice, w}e choose $d$ so that the average time spent by forager bees inside the hive between trips is \dBlue{$\tau_d=0.816$ h} \footnote{\label{foot2}the half-life $\tau$ is derived from the arrival or departure rate, and defined in Eq.~(\ref{eq.tau}) below.} \cite{rodney2020dietary}.
\end{itemize}

Even if the first approach does not lead to a point estimation of the parameters, in practice most solutions can be restricted to relatively narrow intervals by considering physically-based constraints. Some of these constraints are the average mass of food carried by a single bee $m$, the interval of time between trips $\tau$, and the number of trips needed to satisfy food requirements. Details on the computational methods appear in Materials and Methods Sec.-\ref{App.Data} and the codes are available in the repository~\cite{GitHub}.

\subsection{Results from data analysis}\label{sec.results}

In this section we report and discuss the results obtained by applying the model defined in Sec.~\ref{sec.model} to data using the inference methods explained in Sec.~\ref{sec.inference}.

\paragraph{Data.} We \dBlue{used} 24h time series of weights $W'(t)$ sampled every $\delta t=1$ minute for ten different hives on different days. The data was collected between 27th of November 2017 and 2nd of March 2019, from hives located at Macquarie Park, NSW, Australia (33° 46′ 06.6″ S 151° 06′ 43.8″ E, see \cite{colin2021effects}). \dBlue{ Here we present data from the 7\textsuperscript{th} of April in Table 2 and 5 additional dates in Figure 5 (b) and (c). In all cases, the daily maximum temperature was above 16$^{\circ}$C (which is sufficient for honeybees to forage \cite{clarke2018predictive}), and no rain was recorded (ensuring that the recorded weights are not affected by precipitation). This dataset also contains an independent estimation of the total number of bees in each of the hives, obtained from measures of the total weight of the hive with and without bees (performed two days after the day we use to infer the parameters of our model)~\cite{colin2021effects}. As a pre-processing step to the analysis of weight time series,} we removed outliers that correspond to exogenous influences  as described in Materials and Methods Sec.-\ref{App.Computational_methods}. 

\begin{figure}[!ht]
\centering
\includegraphics[scale=0.45,trim=1cm 2cm 1cm 1cm, angle =0 ]{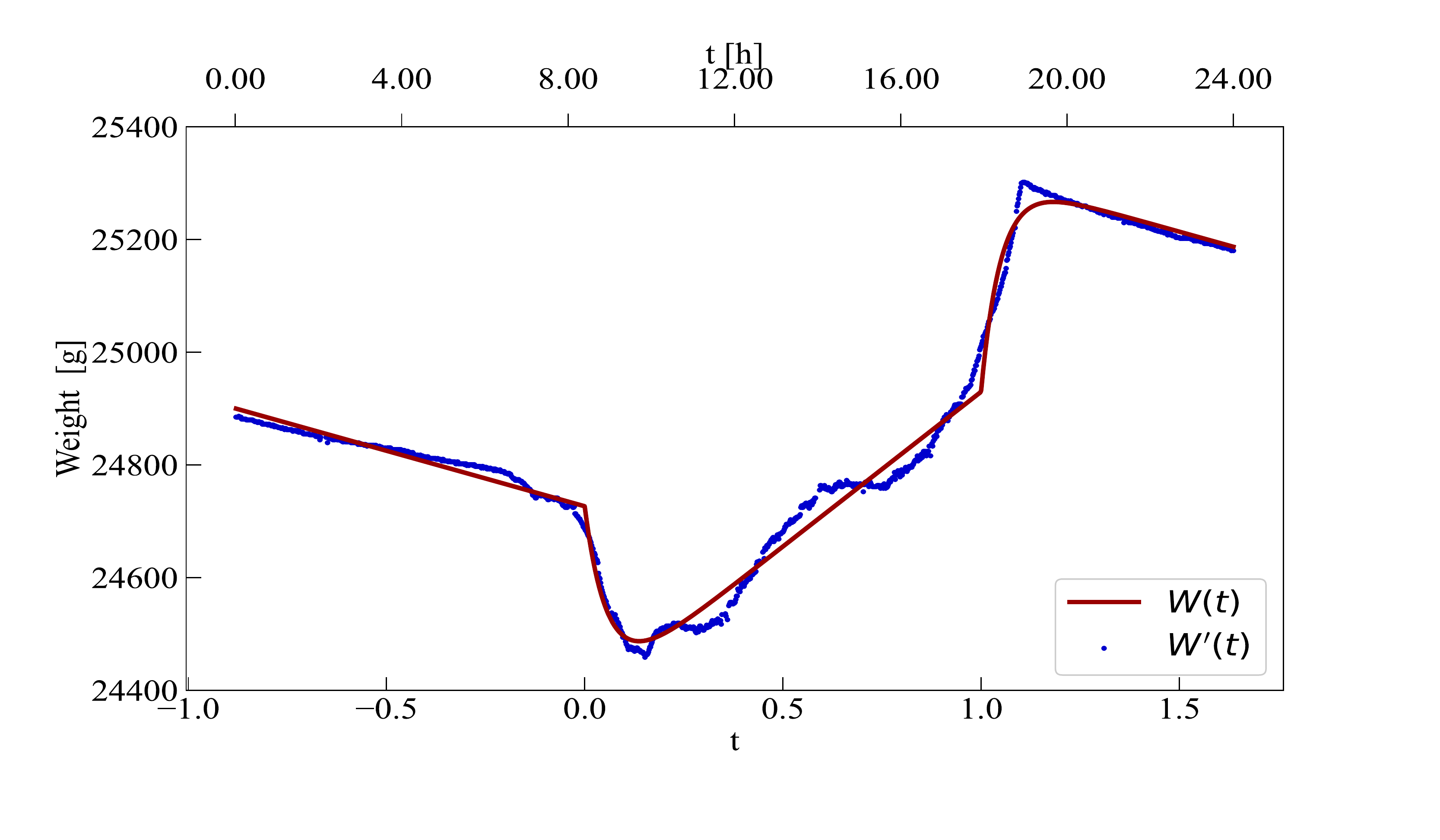}
\caption{Comparison between the model and data. Data corresponds to measurements on 2018-4-7 for `Hive 6'. By fixing $\tau_d = 49$ min, the estimated parameters are $N_{max}=5830 $ bees, $W_{0} =24068\,$g, $m=0.03$ g/bee, $a_{1}=0.99$ $h^{-1}$, $a_{2}=2.1$ $h^{-1}$ , $d=0.85$ $h^{-1}$, $l=20.66$ g/h, $t_{0}=8.40\,$h and $t_{1}=17.93\,$h.} \label{fig.4}
\end{figure}

\paragraph{Representative case.} Fig~\ref{fig.4} shows the comparison between our model with inferred parameters and data for one specific hive. We see how the combination of the three piece-wise continuous curves in our model (for $t<0$, $0\le t \le 1$, and $t>1$, respectively) provides a reasonable account of the characteristic pattern of within-day hive weight variations. The main advantage of our methodology is the mechanistic generative model associated to this curve which ensures the parameters describing the curve have a relevant meaning. For instance, we estimate that on this day there were $N_{max}=5,830$ active forager bees, that they brought on average $0.03$ g of food on each trip, and that trips were on average \dBlue{$\tau_{a_1}=0.70 \, \textrm{h}$} $\textsuperscript{\ref{foot2}}$ long. 
 
 \begin{figure}[!ht]
\centering
\includegraphics[scale=0.33,trim=11cm -.7cm 9cm 1cm, angle =0 ]{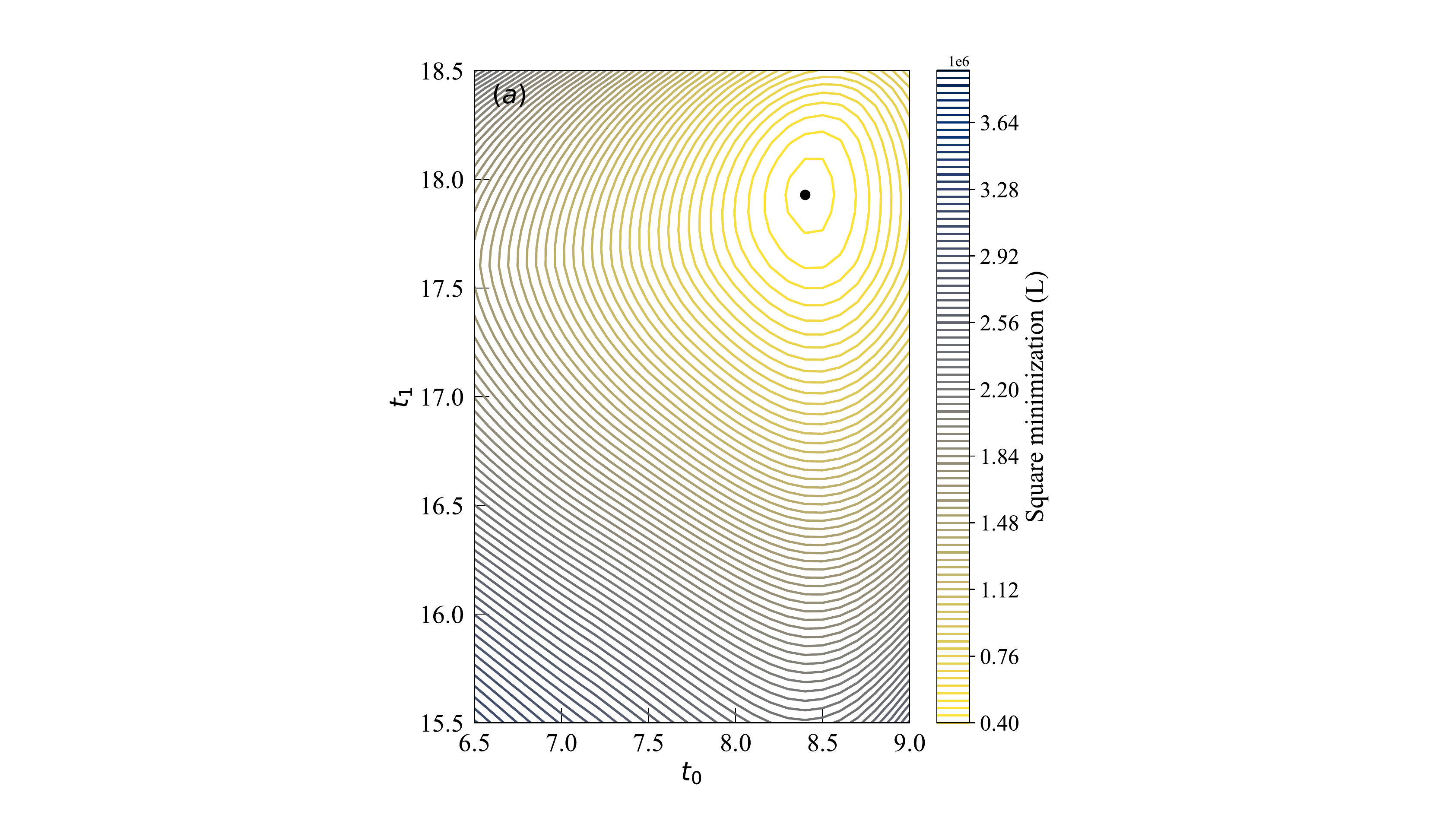}
\includegraphics[scale=0.33,trim=12cm -.7cm 9cm 1cm, angle =0 ]{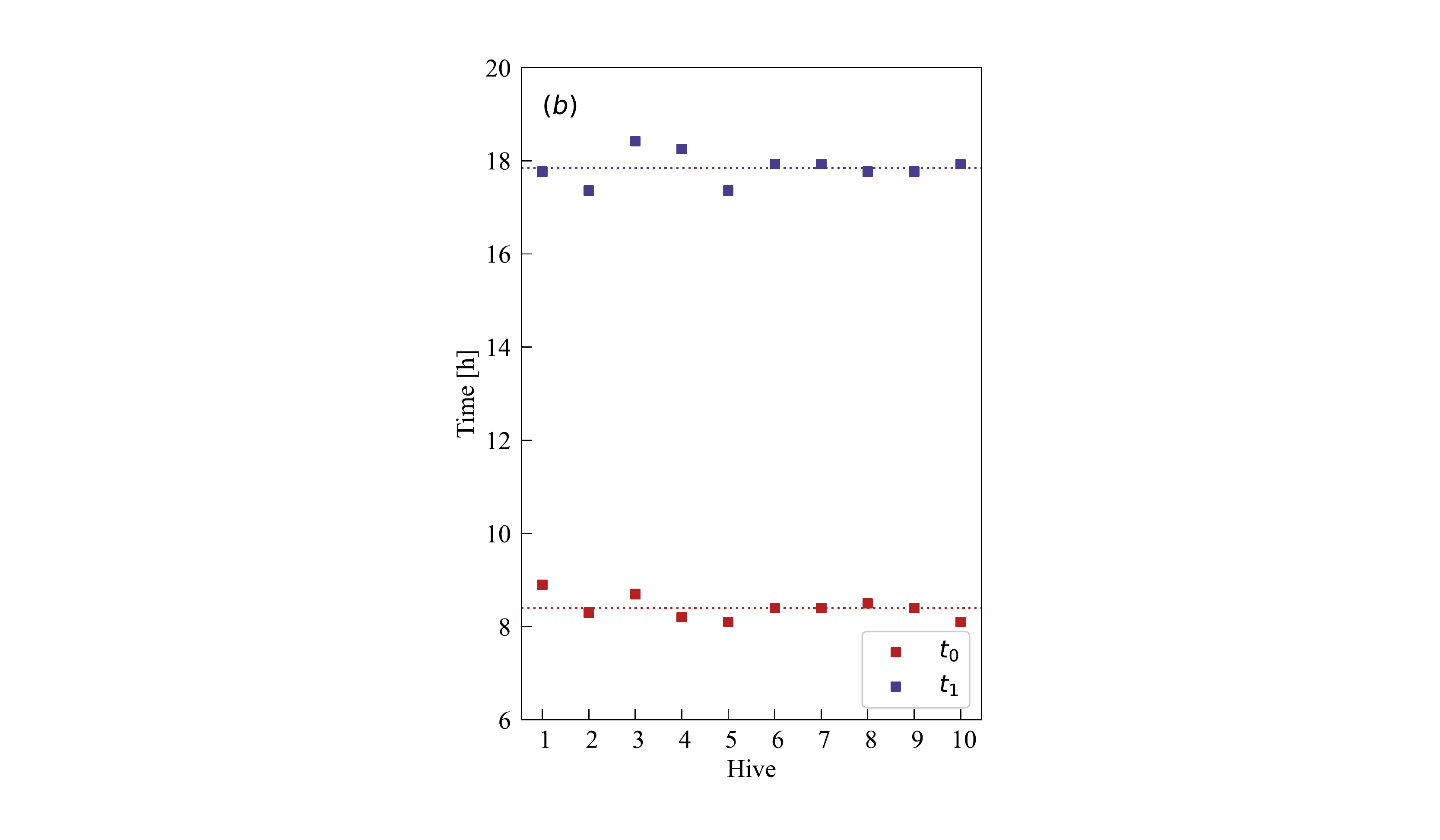}
\includegraphics[scale=0.33,trim=12cm  -1.5cm 9cm 0cm, angle =0 ]{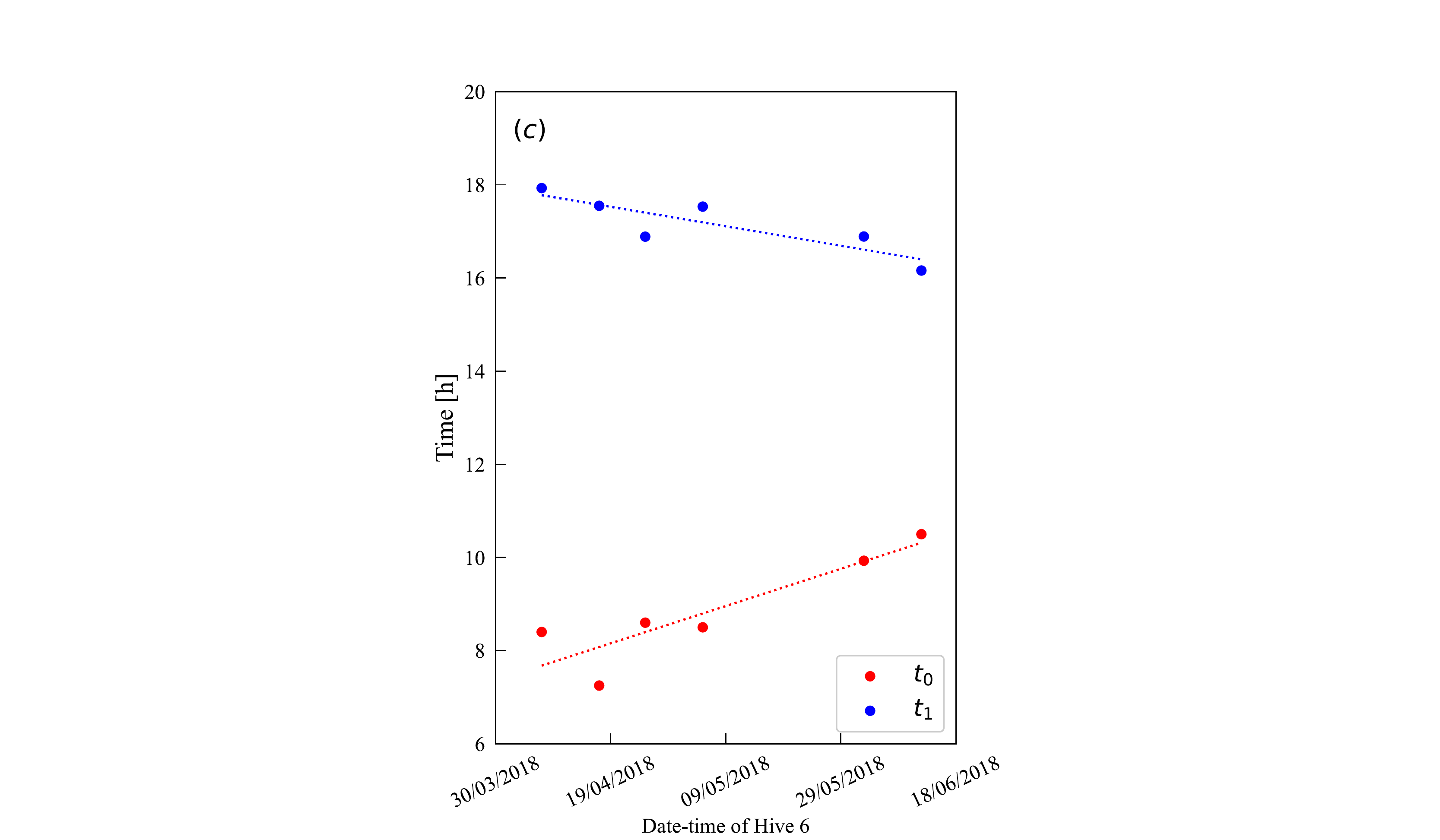}
\caption{Inference of activity times $t_{0}$ and $t_{1}$. (a) Contour plot of $L$ obtained performing multiple minimizations at fixed $t_0$ and $t_1$ (`Hive 6' at 2018-4-7). The minimum $L$ is at $t_{0}=8.40\, \mathrm{h}$  and $t_{1}=17.93 \, \mathrm{h}$. (b) \dBlue{Estimated $t_{0}$ and $t_{1}$ fluctuate around similar values for ten different hives located in close proximity when analysing data sets during the same day. E.g. at 2018-4-7, the different activity times oscillate around $t_{0}=8.4 \, \mathrm{h}$ and $t_{1}=17.85 \, \mathrm{h}$, respectively (dashed lines).} (c) Time evolution for the optimum $t_{0}$ and $t_{1}$ for `Hive 6' only. Transitioning from Autumn to Winter, \dBlue{we observe a linear increase for $t_{0}$ and  a linear decrease for $t_{1}$ (linear regression).}}
\label{fig.ts}
\end{figure}

\paragraph{Activity time.} As a further test of the robustness of our inference method, and the meaningfulness of its outcome, we look in detail at the inferred times in which foragers start ($t_0$) and stop ($t_1$) departing the hive on different days and different hives. The results summarized in Fig.~\ref{fig.ts} show that: (a) a single well-defined optimal parameter value exists, confirming the first and crucial step of our inference procedure; (b) similar values are obtained across multiple hives in the same day and similar location, in agreement with the expectation that the initial times are similar for hives at similar locations; and (c) the estimated values for the same hive varies throughout the year, with the period of activity becoming shorter as the days become shorter (in the southern-hemisphere winter). These results confirm the success of our approach which is to treat these times as additional parameters to be inferred from the data simultaneously with the other parameters.

\begin{table}[!ht]
\small
\begin{tabular}{ |P{1.7cm}|P{1.0cm}|P{1.0cm}|P{1.0cm}|P{1.0cm}|P{1.0cm}|P{1.0cm}|P{1.0cm}|P{1.0cm}|P{1cm}|P{1cm}|   }
 \hline
 \rowcolor{lightgray}\multicolumn{11}{|c|}{Hives} \\
 \cline{1-11}
 \rowcolor{lightgray}$\theta$ & 1 & 2 & 3 & 4 & 5 & 6 & 7& 8 & 9 & 10 \\
  \hline
 \rowcolor{lightgray}\multicolumn{11}{|c|}{Robust estimation}   \\
 \hline
$t_{0}\,$(h)& 8.90  & 8.30   &  8.70   & 8.20   & 8.10   & 8.40  & 8.40  & 8.50 & 8.40  & 8.10 \\
\hline 
$t_{1}\,$(h)& 17.77 & 17.36  & 18.42   & 18.26  & 17.36  & 17.93 & 17.93 & 17.77& 17.77 & 17.93 \\
\hline
 $l\,\,$(g/h)    & 9.762  & 40.805 & 5.600 &  16.628 & 33.183 & 20.663 & 15.016 & 3.974 & 7.687 & 4.399\\
\hline
\dBlue{$A$(g)}& 17,503& 23,384  & 17,878 & 26,809 & 23,412 & 24,731 &  21,244 & 18,057 & 19,485 & 19,364\\
\hline
\dBlue{$B$(g)}& 17,409 & 22,.849  & 17,808 & 26,562 & 22,905 & 24,369 &  20,940 & 17,925 & 19,342 & 19,262\\
\hline
$\alpha \,$(g/h)     & 16.594  & 95.081  & 4.595 & 38.529 & 72.577  & 59.560 &  24.780 & 4.721 & 23.754 & 4.476\\
\hline
$t_{c} \,$(h)     & 9.521  & 9.598  & 9.852 & 9.633 & 9.565  & 9.706 &  10.525 & 13.485 & 9.953 & 11.423\\
\hline
\rowcolor{lightgray}\multicolumn{11}{|c|}{Interval estimation}   \\
\hline
$N_{max}\,$ (bees)  & [1950, 5400]  & [5050,   8200] & $\,$[940, $\,\,$  2500] & [2310, 3450] & [7180, 7650] & [4150, 6030] & $\emptyset$ & $\,$ $\emptyset$ & $\emptyset$ & $\,$[840, $\,\,$  940]\\
\hline
$W_{0}$(g)  & [16892, 17282] &[22451, 22794]&[17596, 17772] &[26413, 26537] &[22542, 22594] &[24045, 24254] & $\emptyset$ &$\emptyset$ & $\emptyset$ &[19257, 19267]\\
\hline
$a_{1}$(1/h)&[3.59, 4.79] &[0.61, 0.92]&[1.44, 2.63]&[0.54, 0.79]&[0.81, 0.85]&[0.76, 1.01]& $\emptyset$ &$\emptyset$ & $\emptyset$ &[0.20, 0.25]\\\hdashline
$\tau_{a1}$(h)&[0.14, 0.19] &[0.75, 1.13]&[0.26, 0.48]&[0.88, 1.28]&[0.82, 0.86]&[0.69, 0.92]& $\emptyset$ &$\emptyset$ & $\emptyset$ &[2.75, 3.40]\\
\hline
$a_{2}$(1/h)&[1.34, 1.34]&[1.01, 1.02]&[3.20, 3.21]&[2.43, 2.47]&[1.07, 1.07]&[2.10, 2.11]& $\emptyset$ &$\emptyset$& $\emptyset$ &[2.36, 2.35]\\\hdashline
$\tau_{a2}$(h)&[0.52, 0.52]&[0.68, 0.69]&[0.22, 0.22]&[0.28, 0.29]&[0.65, 0.65]&[0.33, 0.33]& $\emptyset$ &$\emptyset$& $\emptyset$ &[0.29, 0.30]\\
\hline
$d$(1/h)& [0.82, 2.38] & [0.82, 1.41] & [0.82, 2.28] & [0.82, 1.26] & [0.82, 0.87] & [0.82, 1.24] & $\emptyset$ &$\emptyset$& $\emptyset$ & [0.82, 0.94]\\\hdashline
$\tau_{d}$(h)& [0.29, 0.85] & [0.49, 0.85] & [0.30, 0.85] & [0.55, 0.85] & [0.79, 0.85] & [0.56, 0.85] & $\emptyset$ &$\emptyset$& $\emptyset$ & [0.74, 0.85]\\
\hline
\dBlue{$m\left(\frac{\times10^{-3} g}{bee}\right)$} & [7.00, 10.00] & [37.40, 60.00] & [7.00, 12.00] & [38.50, 59.80] & [32.40, 34.10] & [28.80, 39.60] & $\emptyset$ & $\emptyset$ & $\emptyset$ & [39.00, 45.00]\\
\hline
\rowcolor{lightgray}\multicolumn{11}{|c|}{Estimation assuming \dBlue{$\tau_d=0.816$ h} }   \\
\hline
$N_{max}\,$ (bees)  & 5220$\pm$0 & 7930$\pm$0  & 2410$\pm$0  & 3330$\pm$0 & 7400$\pm$0  & 5830$\pm$0 &  3540$\pm$0 & 970$\pm$0 & 1260$\pm$0 & 910$\pm$0\\
\hline
$W_{0}$(g)  & 16912$\pm$3 & 22481$\pm$7 & 17606$\pm$5 & 26426$\pm$3 & 22570$\pm$4 & 24068$\pm$5 & 20840$\pm$1 & 17946$\pm$4& 19339$\pm$2& 19260$\pm$1\\
\hline
$a_{1}$(1/h)& 4.77$\pm$0.6  & 0.90$\pm$0.0 &  2.60$\pm$0.2 &  0.77$\pm$0.0  & 0.83$\pm$0.0  & 0.99$\pm$0.1 & 0.50$\pm$0.0 &  0.11$\pm$0.0 & 0.44$\pm$0.0 & 0.24$\pm$0.0\\
\hline
$a_{2}$(1/h)     & 1.34$\pm$0.1  & 1.02$\pm$0.1 &  3.21$\pm$0.3  & 2.47$\pm$0.2  & 1.07$\pm$0.1  & 2.11$\pm$0.1 & 2.00$\pm$0.3 &  1.56$\pm$0.1  & 1.84$\pm$0.1 & 2.35$\pm$0.2\\
\hline
$d$(1/h)     & 0.85$\pm$0.1  & 0.85$\pm$0.0 &  0.85$\pm$0.1  & 0.85$\pm$0.1  & 0.85$\pm$0.0  & 0.85$\pm$0.1 & 0.85$\pm$0.0 &  0.85$\pm$0.0  & 0.85$\pm$0.1 & 0.85$\pm$0.0\\
\hline
\dBlue{$m\left(\frac{\times10^{-3} g}{bee}\right)$}& 10.0$\pm$1 & 38.0$\pm$2 &  10.0$\pm$1  & 40.0$\pm$3  & 33.0$\pm$1  & 30.0$\pm$3 & 33.0$\pm$1 &  53.0$\pm$2  & 77.0$\pm$3 & 41.0$\pm$2\\
 \hline
\end{tabular}
\caption{Inferred parameters for 10 different hives on the same day (2018-4-7). Values $\emptyset$ in the interval estimation indicates that there was no interval of parameters satisfying all imposed constraints for this hive.The values of $\tau$ next to the parameters $a_1,a_2,$ and $d$ correspond to the time spent outside (respectively, inside) the hive and were computed using Eq.~(\ref{eq.tau}). The uncertainties in the values of the estimation assuming \dBlue{$\tau_d=0.816$ h} minutes, reported under the curly brackets, were computed using bootstrapping, see Materials and Methods Sec.-\ref{App.boostrapping}. } 
    \label{tab.inferred}
\end{table}

\begin{figure}[ht!]
\centering
\includegraphics[scale=0.4,trim=10cm .5cm 8cm 1cm, angle =0 ]{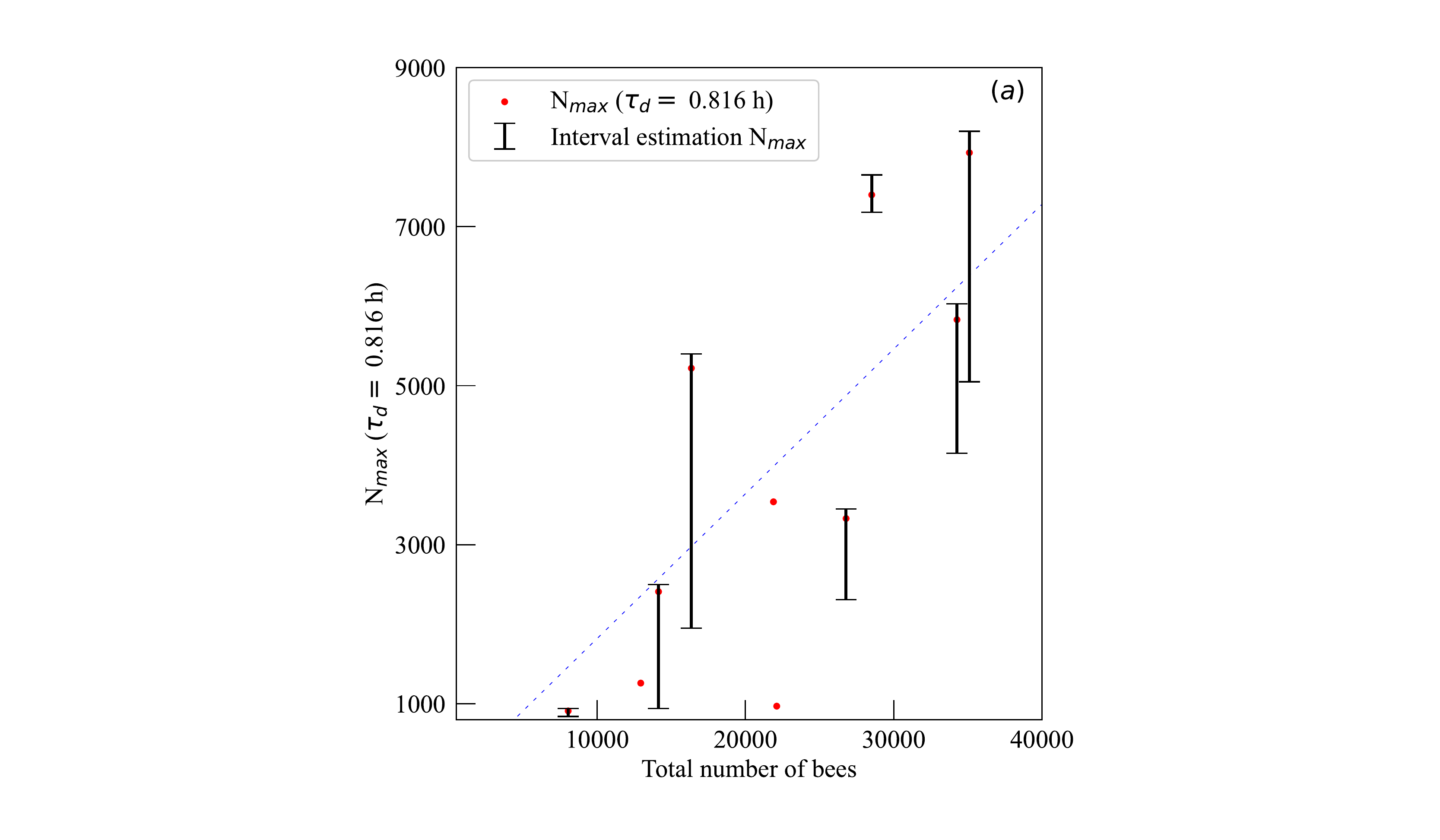}
\includegraphics[scale=0.4,trim=9cm 0cm 8cm 4cm, angle =0 ]{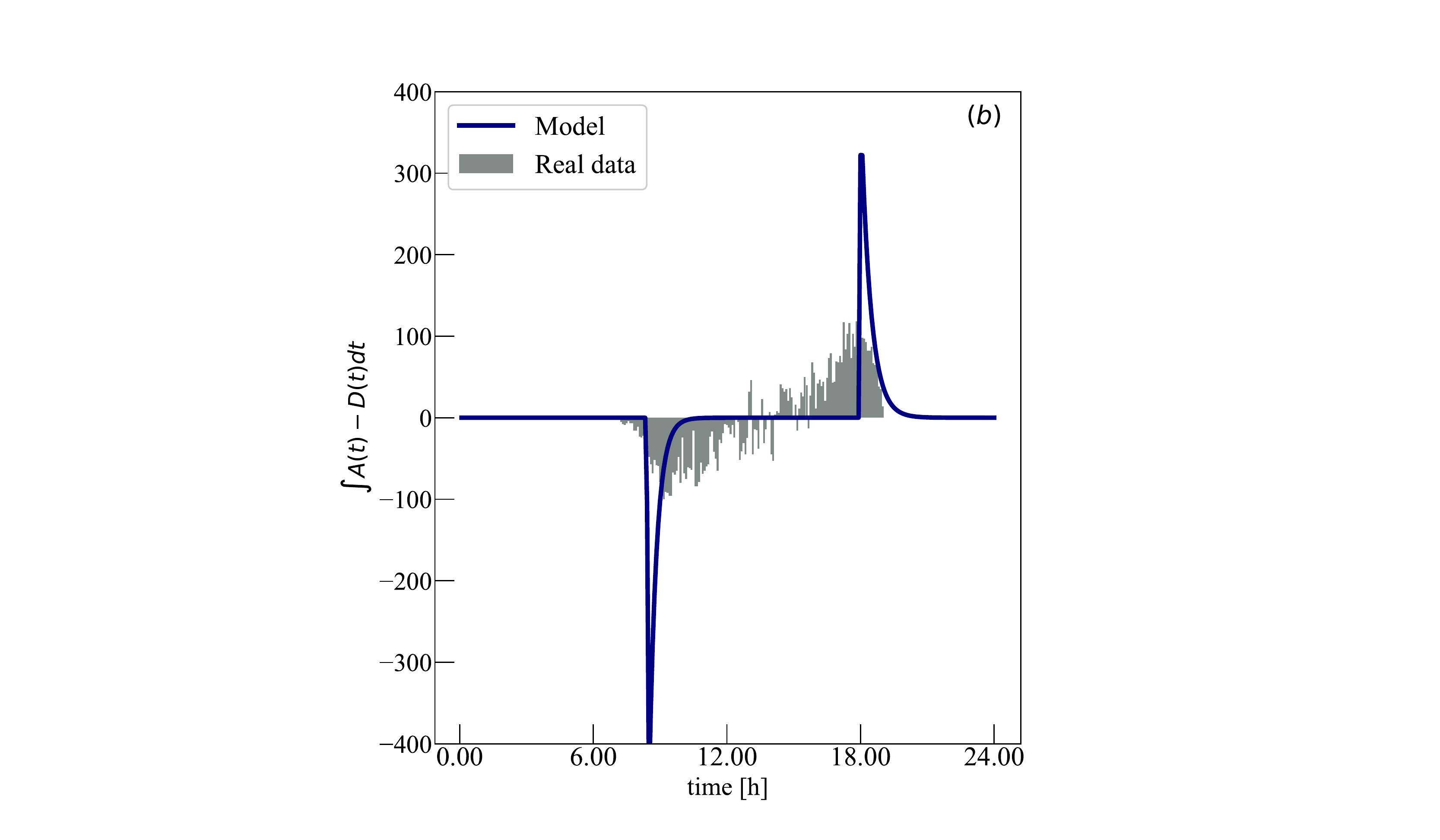}
\caption{\dBlue{Comparing our model to independent measurements. (a) Our estimation of the number of active foragers $N_{max}$ correlates positively with the total number of bees ($R^2=0.596, p=0.036)$. The dashed straight line corresponds to 18.2\% of all bees being foragers, constant across all hives.  The total number of bees was estimated through an independent measurement performed on 2018-04-9, two days after the data used in our inferrence~\cite{colin2021effects}. }(b)The plot represents the difference between the number of bees arriving and bees departing a honey-bee hive during a 5 minutes interval. \dBlue{The model prediction (blue line) was computed integrating $A(t)-D(t)$ over each time interval. The real data was measured using bee-tracking methods.} }\label{fig.ad} 
\end{figure}

 \paragraph{Systematic analysis.} Motivated by the success of our methodology in the cases above, we applied it systematically to 10 different hives on the same day (2018-4-7). The results obtained from our two inference methods are reported in Table \ref{tab.inferred}. 
 The comparison of the inferred parameters of these hives with expectations about the behaviour of bees and hives provide further evidence that our mathematical model is capturing meaningful information from the data. In particular:

\begin{itemize}
    \item The time forager bees start ($t_0$) and stop ($t_1$) departing the hives are aligned with sunlight. 
    \item The time of minimum weight was inferred to be   around $t_c=9:45$ a.m. shortly after $t_0$ (for all  hives except `Hive 8', which presents a later minimum). This result is compatible with the minimum daily weights reported in experimental data \cite{hambleton1925effect}, which reveal minimum weights around $9:00$ a.m. Recently, results from segmented linear regression showed minimum losses can take place even 4 hours after sunrise \cite{holst2018breakfast}. 

    \item The estimation of the number of forager bees $N_{max}$ is around a few thousand bees, in line with alternative methods of estimation. \dBlue{In particular, Figure~\ref{fig.ad}(a) shows that the total number of bees per hive correlates positively with our estimation of $N_{max}$ ($R^2=0.596$, p=0.036)\footnote{\dBlue{In all our analysis of correlation between variables, the reported p-value p is the probability of obtaining a positive coefficient of determination $R^2$ equal or larger than the reported $R^2$ under the null hypothesis that the variables are independent from each other.}}. The fraction of forager bees (assuming $\tau_d=0.816h$) is estimated to be around $18\%$, comparable to previous estimations (around $25\%-30\%$ in Refs. \cite{van2015foraging,rodney2020probabilistic}). Additionally,} we obtained a  positive correlation between $N_{max}$ and \dBlue{rate of weight gained during the day $\alpha$, ($R^2$=0.778, p=0.0056)} consistent with the previously reported link between the size of the population and the amplitude of weight variations during the active foraging period \cite{holst2018breakfast}.
    \item The average amount of food brought back to the hive by a bee was inferred to be typically in the range \dBlue{$m \in [10,40] \times 10^{-3}$ g/bee}, which is comparable to reported crop loads carried by active foraging bees when returning to the hive with loads of pollen \cite{garcia2004variations}, water \cite{visscher1996honey}, and different sweet solutions \cite{afik2006effect}. The two exceptions (Hives 8 and 9) show values up to \dBlue{$m=77 \times 10^{-3} $ g/bee} and possible correspond to sugar concentrations with total dissolved solids (TDS) of 60\% \cite{afik2006effect}. 
    
    \item In 90\% of the cases, $a_1<a_2$, which is in line with the expectation that bees return faster to the hive at dusk (after they stop leaving).
    
    \item The typical time spent by forager bees inside and outside the hive is computed through an analogy with the half-life, $\tau$, which represents the time that half of the foragers' colony will need to go outside the hive or vice-versa. The half-life, $\tau$ depends on the departure or arrival rate $r \in \{d,a_1,a_2\}$ in hours as
    \begin{equation}\label{eq.tau}
         \tau = (t_1-t_0) \frac{\ln 2}{r}.
    \end{equation}
    From the calculated values of $d,a_{1,2},$ and $t_1-t_0$, we determine that the time in the hive between trips, $\tau_d$, ranges from 0.29h to 0.85h and the typical time of a foraging trip,$\tau_a$, is in between 0.14h to 3.40h. These estimations align with the results in Refs.~\cite{colin2022evaluating,rodney2020dietary,rodney2020probabilistic}. The only hive that shows values up to 6.25 h for trip duration is Hive 8, this hive shows behaviour outside the ones predicted by our model, as evidenced by $\emptyset$ in the interval estimation. However, some reports points that bees are able to be outside the hive 6.25 h at the age of 36 days \cite{he2013assessment}. 
\end{itemize}

\paragraph{Arrivals and departures.} Finally, we test the predictions of our model against independent experimental measures that tracked the number of bees that arrive and depart from a hive at a time interval~\cite{colin2022evaluating}. Figure~\ref{fig.ad}(b)  compares these measurements to the assumption in our model and reveals the limitations of our simplifying choice of constant and piece-wise linear arrivals and departures in Eq.~(\ref{eq.as}). \dBlue{While our model captures the main characteristics of the data, unsurprisingly, bee activities do not start or end abruptly and the measured data shows smoother variations than the prediction of our model. A similar behaviour can be observed for $W(t)$} -- such as in Fig.~\ref{fig.3} --  where the measured data do not show the abrupt changes seen in the model curve at $t=0$ and $t=1$ (discontinuous derivative). This limitation can be partially overcome within our modeling framework --  Eqs.~(\ref{eq.W}) to~(\ref{eq.F}) --  by changing the assumptions of constant $a(t)$ and $d(t)$ -- e.g., in Eq.~(\ref{eq.as}). For instance, the next simplest alternative is to consider that the arrival remains constant $a(t)=a$ but that the departures varies continuously from $0$ at time $t=0$ to a maximum value $d_{max}$ at time $t=0.5$ and back to $0$ at $t=1$. The simplest polynomial with these characteristics is 
\begin{equation}\label{eq.dmax}
    d(t) = 4 d_{max} t (1-t).
\end{equation}
In this case, Eq.~(\ref{eq.x}) is a linear first-order equation and the solution is:
\begin{equation}\label{eq.dmaxSol}
    x(t) = \frac{C}{e^{r(t)}} + a \int e^{r(t) dt} dt,
\end{equation}
where $r(t)=at+2d_{max} t^2 - \frac{4}{3} d_{max} t^3$ and $C$ is the arbitrary constant to be fixed by the initial condition $x(0)=1$. This alternative model has the same number of parameters ($d_{max}$ instead of $d$), but requires additional computational resources to numerically find the solution for $x(t)$ through the integration of Eq.~(\ref{eq.dmaxSol}). This example illustrate the flexibility of our model to consider more realistic settings and also the trade-off between model complexity and the need of computational approaches.

\section{Discussion}\label{sec.conclusions}

We introduced a simple mathematical model that describes the hive processes affecting within-day weight variations. Simplistic assumptions of the departure and arrival of bees allow us to obtain closed form solutions for the model that result in a continuous, piece-wise differentiable, curve $W(t)$ with $10$ biologically-relevant parameters. We inferred the parameters of $W(t)$ for $10$ different hives and for different days across the year. Overall, the results reveal that the parameters we infer are compatible with previous knowledge of bee behavior.

The main advantage of our approach when compared to previous works is that it combines mechanistic mathematical models with data analysis to obtain interpretable quantitative information about the condition of the hive. Additional advantages of our model approach include: (i) indications from the inference of when results cannot be trusted (either because they are not capturing meaningful information or because optimal results lie outside of the allowed error range); (ii) that our framework allows for the proposal of more sophisticated models, e.g., by considering additional terms contributing to the food dynamics in Eq.~(\ref{eq.F}) or more detailed modes of arrival $a(t)$ and departure $d(t)$ such as in Eq.~(\ref{eq.dmax}). An important methodological expansion of our work would be to consider and compare different type of functional forms in our models. This will require considering cases in which solutions are not obtained in closed forms and the implementation of a Bayesian-model comparison that can compare models of different complexities, in line with modern simulation-based and Bayesian inference that are increasingly applied to dynamical systems of biological interest~\cite{cranmer_frontier_2020,roda_bayesian_2020,barnes2011bayesian}.

The success of our model in extracting meaningful information about bee hives opens the possibility of using it for a systematic analysis of the health of hives across different hives and across time. The following quantities provided by our model are of particular interest:
\begin{itemize}
\item The number of active foragers in a bee colony $N$ is a crucial parameter of the capacity of a colony at sustaining itself. Environmental stressors such as pesticides can cause a rapid depletion of the population of active foragers \cite{colin2019traces}. Estimations of $N$ provide a rapid assessment of the daily foragers population and allow quick intervention and stress reduction to prevent colony collapse \cite{perry2015rapid}.
\item The time spent by foragers outside the hive~$\tau_a$ can be connected to the efficiency of the colony because  long times outside the hive lead to a reduction of the life expectancy of foragers \cite{colin2022evaluating}.
\item The amount of food collected by bees in each trip $m$ allows for a better assessment of foraging difficulty and thus of environments that can best support beekeeping operations~\cite{quinlan2022grassy}.
\item The rate of weight grow of a hive $\alpha$ gives a reliable indicator of whether foragers are supplying their colony properly (leading to optimal colony development) or if instead the colony is endangered (suggesting the existence of stressors or environmental limitations). 
\end{itemize}
Tracking the daily evolution of these parameters opens the perspective of developing weight-based early-warning indicators of bee colony failure that could lead to new strategies for risk control.


\input{Methods}

\section*{Supporting Information}

    \paragraph{Figure 7:} Multiple solutions are obtained from the Hive 6 on $2018-4-7$, and then reduced to a confidence interval.
    \paragraph{Figure 8:} Example of outlier detection for the Hive 10 on $2018-4-7$
    \paragraph{Figure 9:} Probability Density Function for the parameters displayed at Table \ref{tab.parameters} after the bootstrapping process of the Hive 6 on $2018-4-7$.
    \paragraph{Data reporting:} \dBlue{Data is available at the \href{https://link.springer.com/article/10.1007/s13592-020-00837-3}{Supplementary Information} of Ref.~\cite{colin2021effects} and in our repository~\cite{GitHub}.}

    \dBlue{\paragraph{Codes:} The implementation of the proposed model and data analysis methods presented in this paper are available in the GitHub repository (\href{https://github.com/Karina-AriasCalluari/weight_beehive}{Click Here}), and in the Zenodo repository (\href{https://doi.org/10.5281/zenodo.7272480}{Click Here})~\cite{GitHub}.}

\clearpage   
\begin{figure}[!ht]
\centering
\includegraphics[scale=0.37,trim=13cm 4cm 4cm 0cm, angle =0 ]{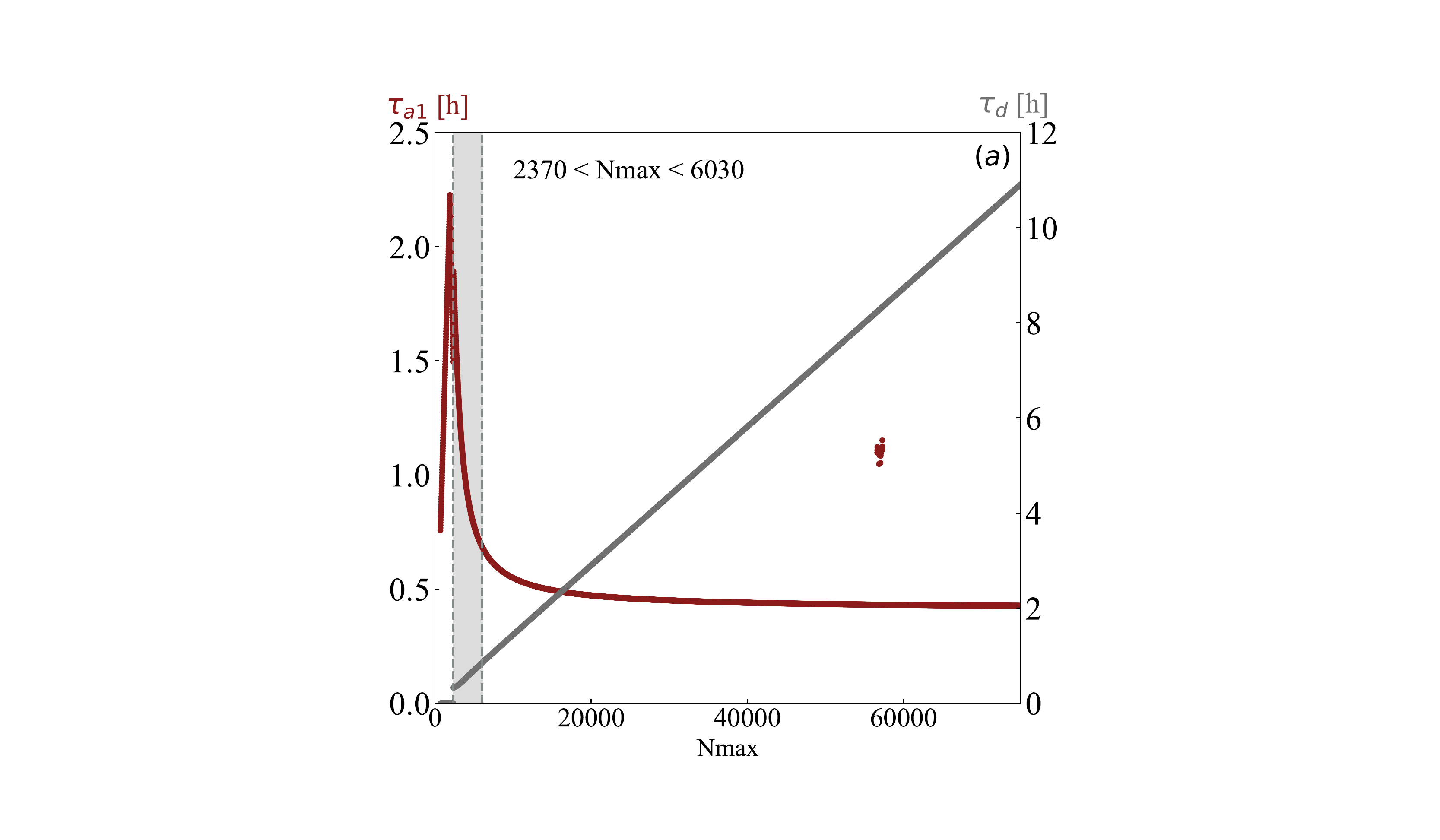}
\includegraphics[scale=0.37,trim=11cm 4cm 12cm 0cm, angle =0 ]{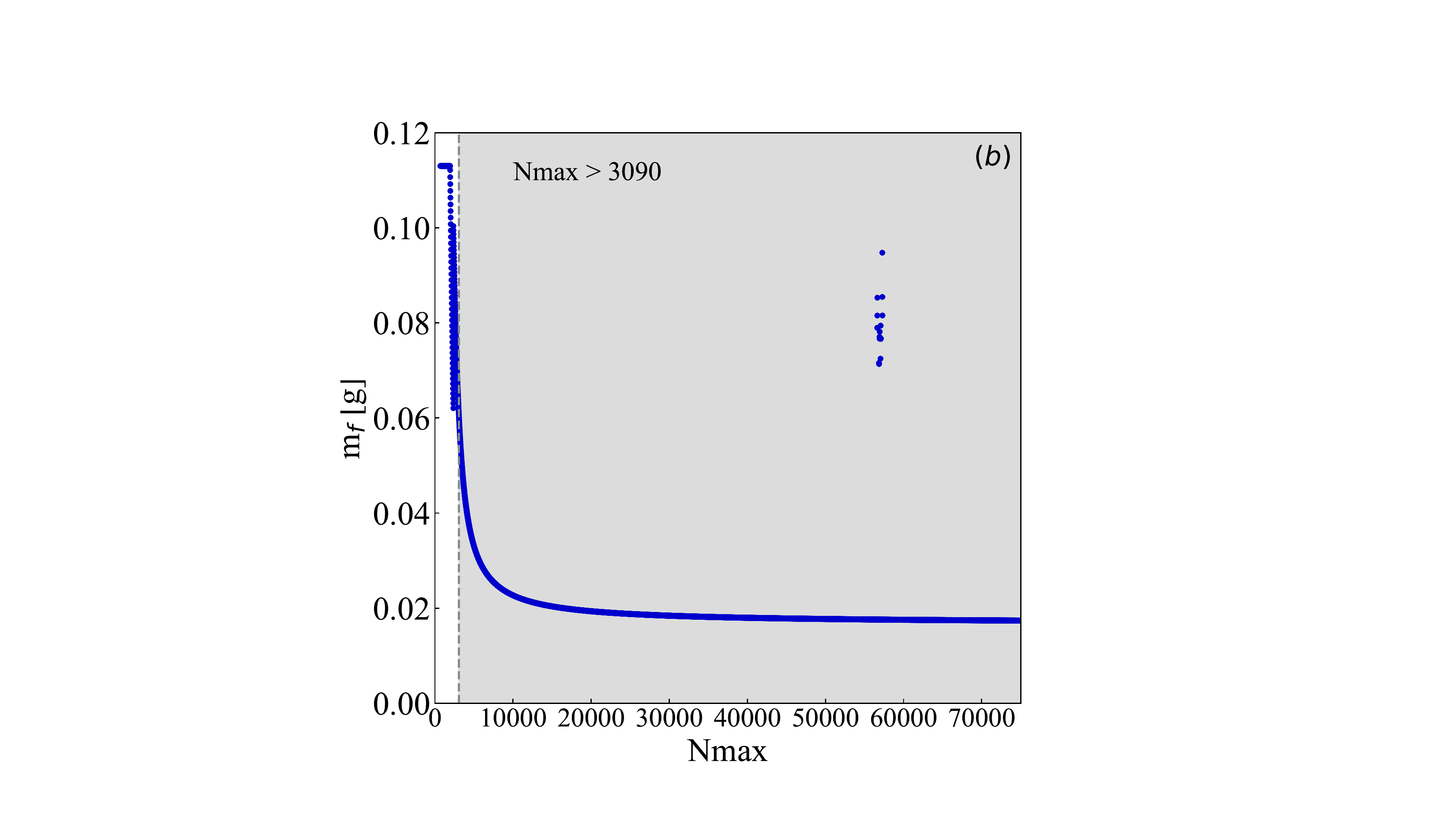}
\includegraphics[scale=0.32,trim=13cm 0cm 4cm 3cm, angle =0 ]{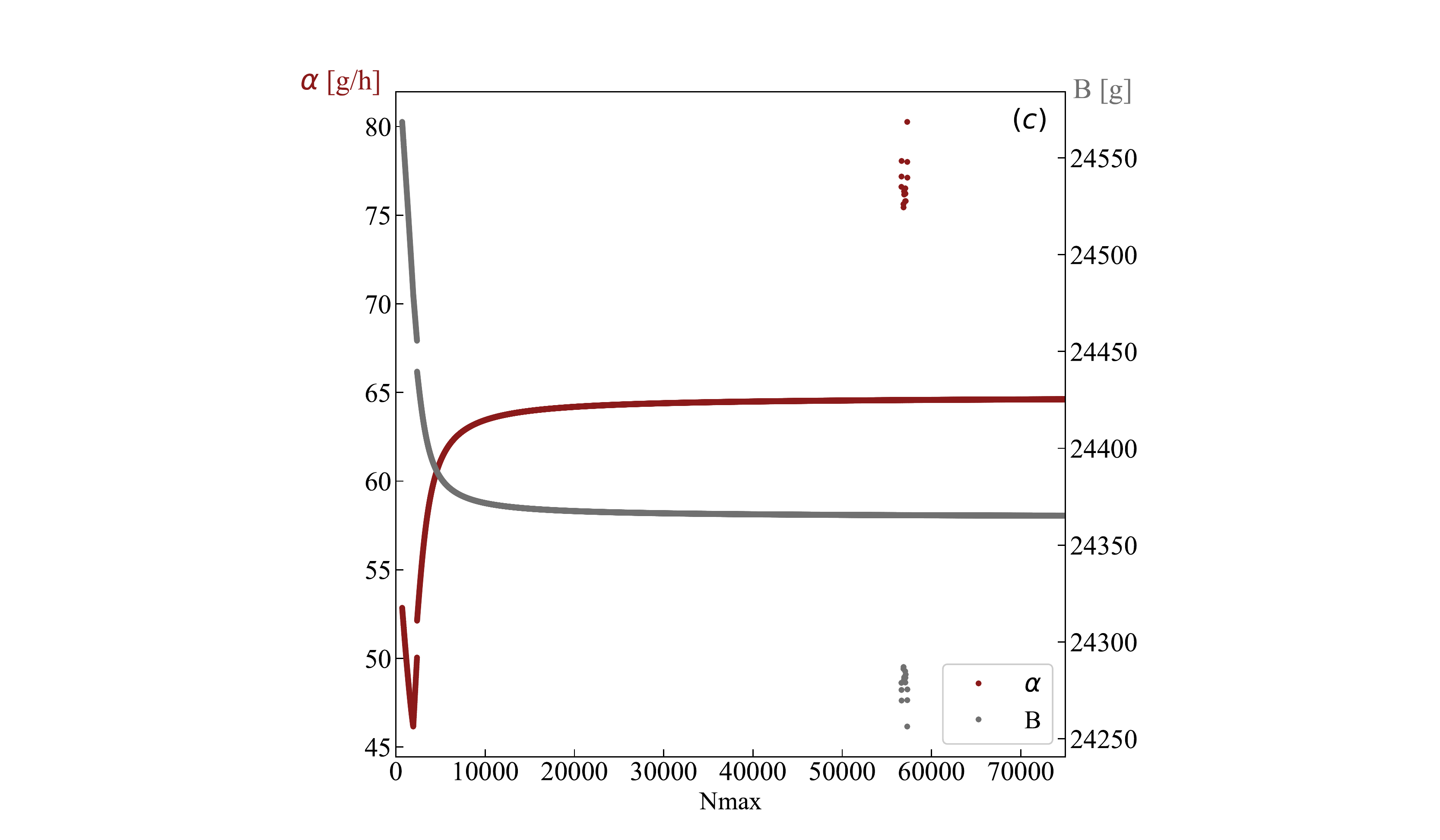}
\includegraphics[scale=0.37,trim=9cm  1cm 12cm -5cm, angle =0 ]{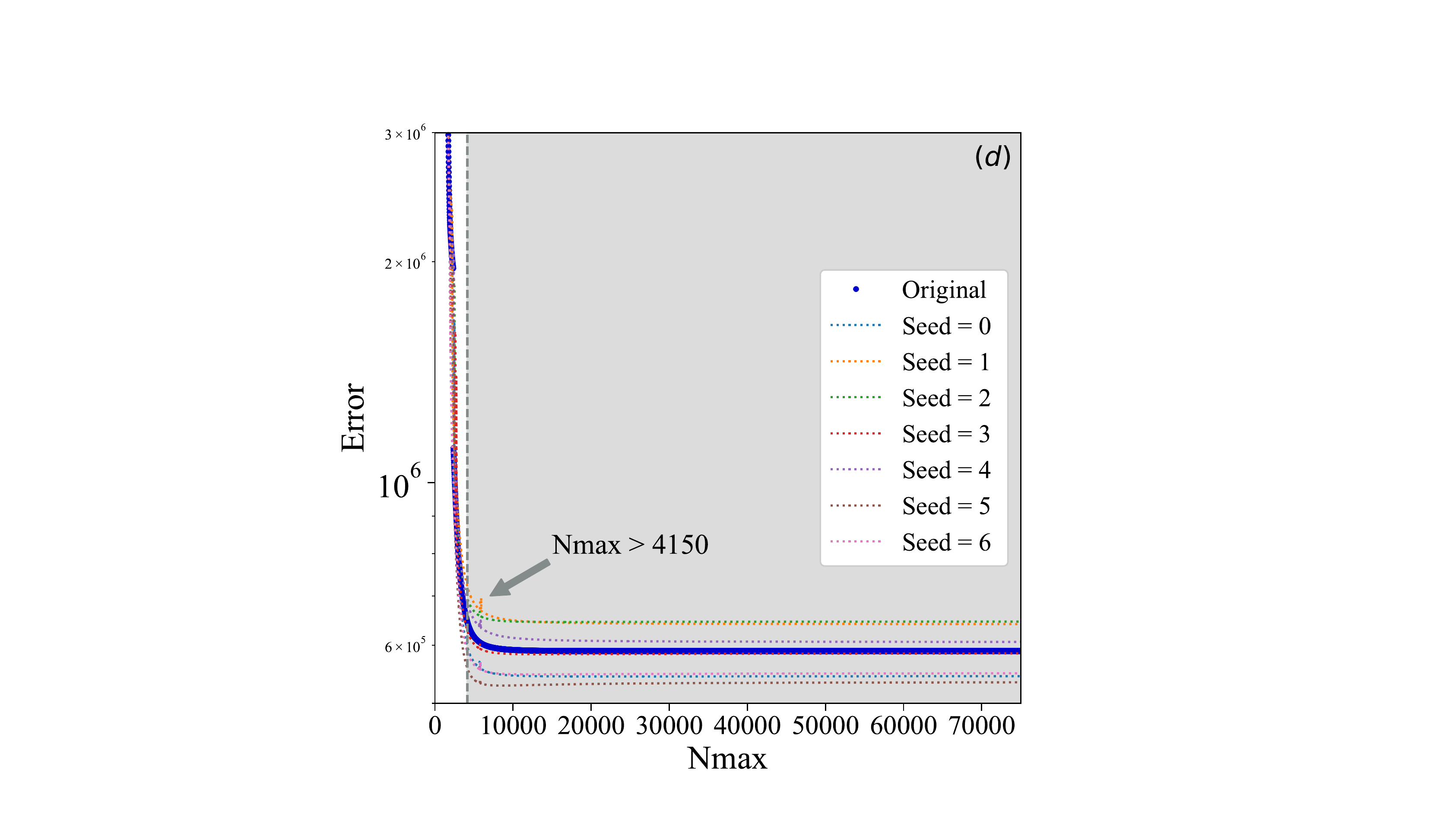}
\caption{Results obtained from the `Hive 6' the day $2018-4-7$ (a) The time spent inside, $\tau_d$, and outside the hive during the day, $\tau_{a_1}$ is compared with reported intervals of time inside the hive taking by the forager bees between trips, and an average of time  outside the hive per flying. Then, a valid interval is selected considering previous reports. (b) The maximum reported food load collected per trip (crop load) is $0.077$ g. The shaded area represents a region where an acceptable crop load is obtained. (c) Representation of two characteristic parameters, the rate of gained weight and the intersection with $t=0$, as $N_{max}$ decreases $B$ and $\alpha$ become unstable. (d) Error calculated as the sum of the squared difference between the model and the real data. The shaded area represents a region where an acceptable error occurs based on bootstrapping simulation. } \label{fig.6}
\end{figure}

\begin{figure}[!ht]
\centering
\includegraphics[scale=0.55,trim=1cm 0cm 1cm 1cm, angle =0 ]{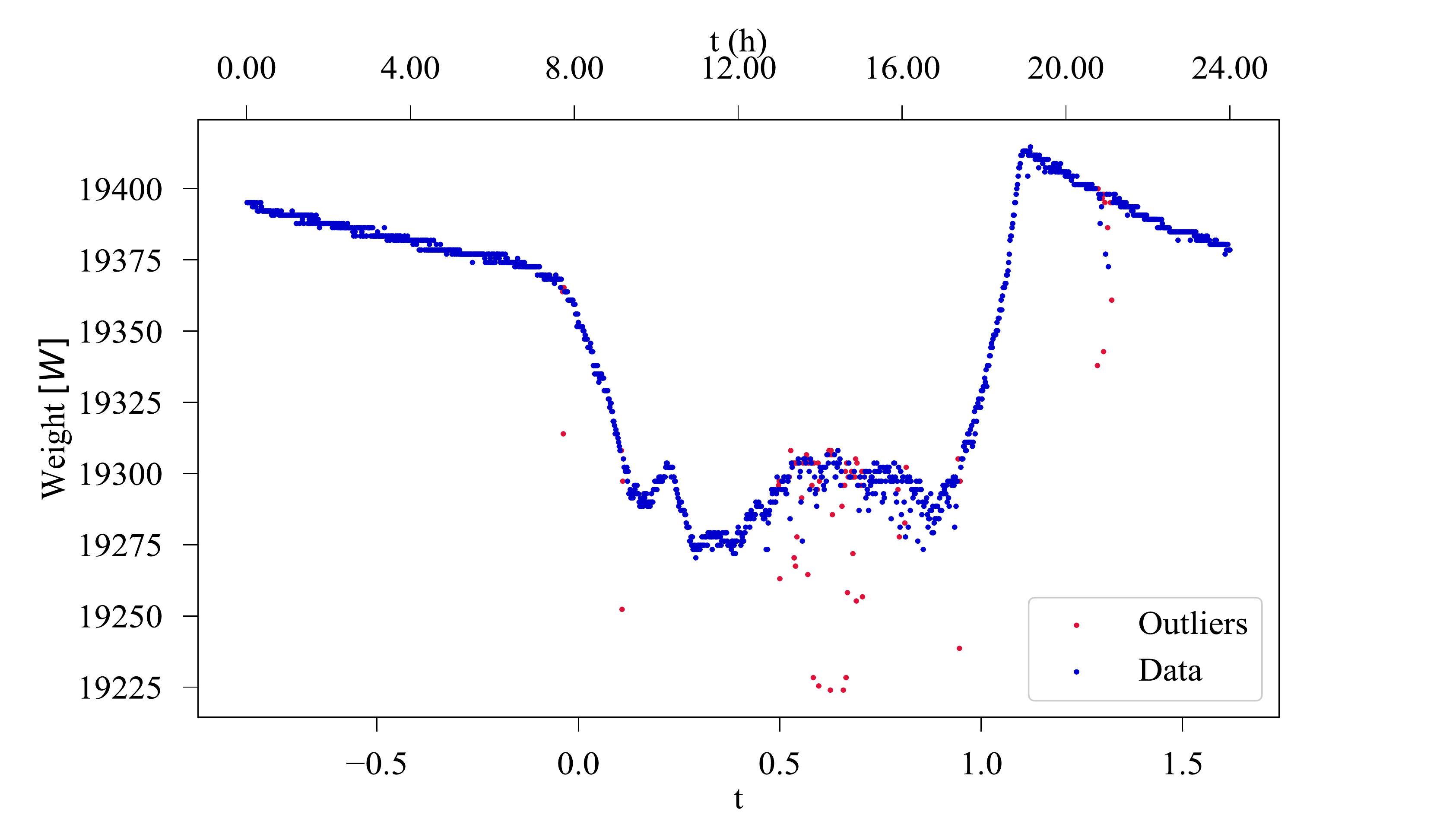}
\caption{Example of outlier detection for the `Hive 10' on 2018 − 4 − 7. The outliers detected are in red color and were removed before the data analysis.} 
\label{fig.outliers}
\end{figure}

\begin{figure*}[!ht]
\centering
\includegraphics[scale=0.275,trim=11cm -2.5cm 5cm 1cm, angle =0 ]{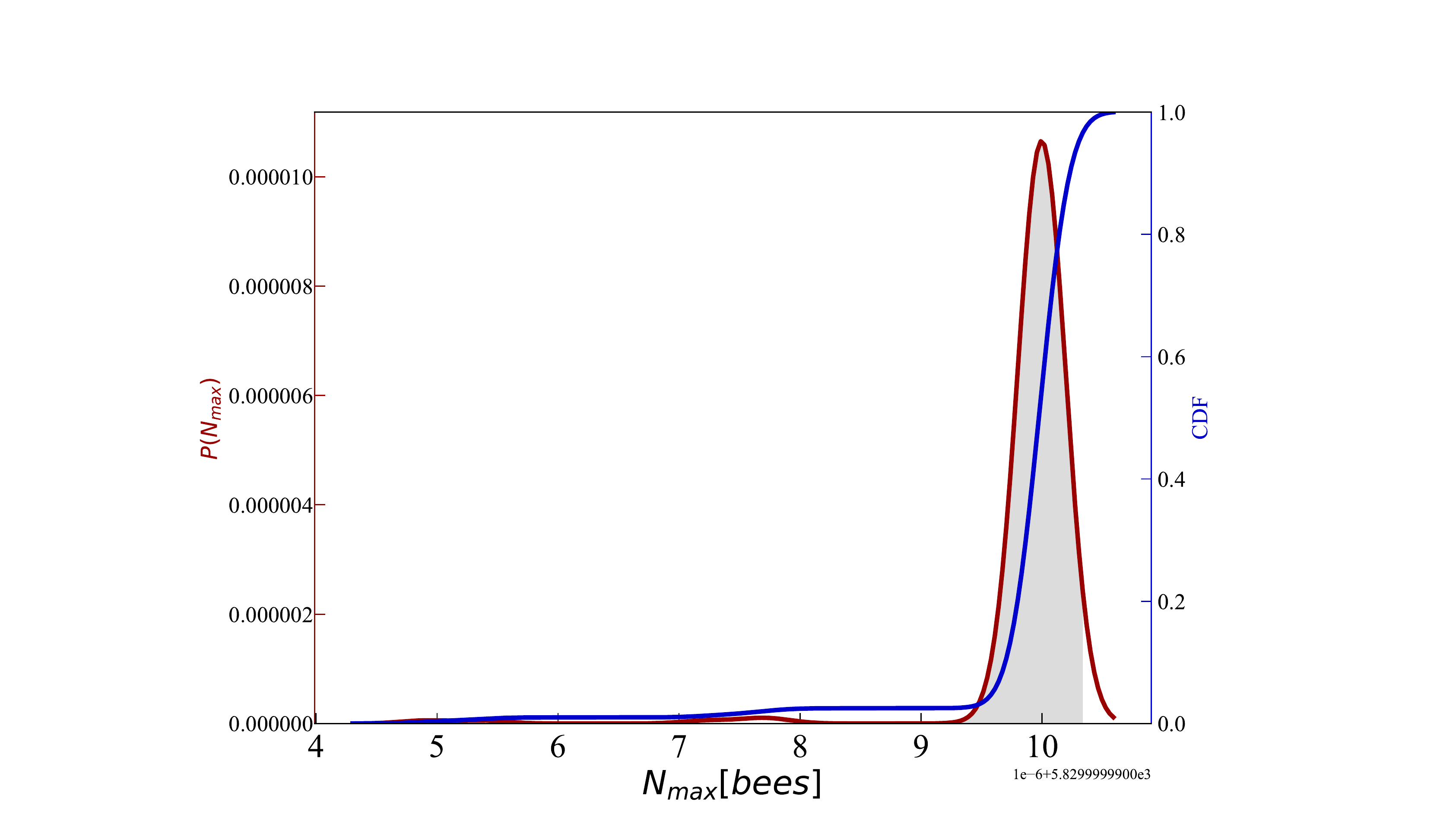}
\includegraphics[scale=0.275,trim=4cm -2.5cm 8cm 1cm, angle =0]{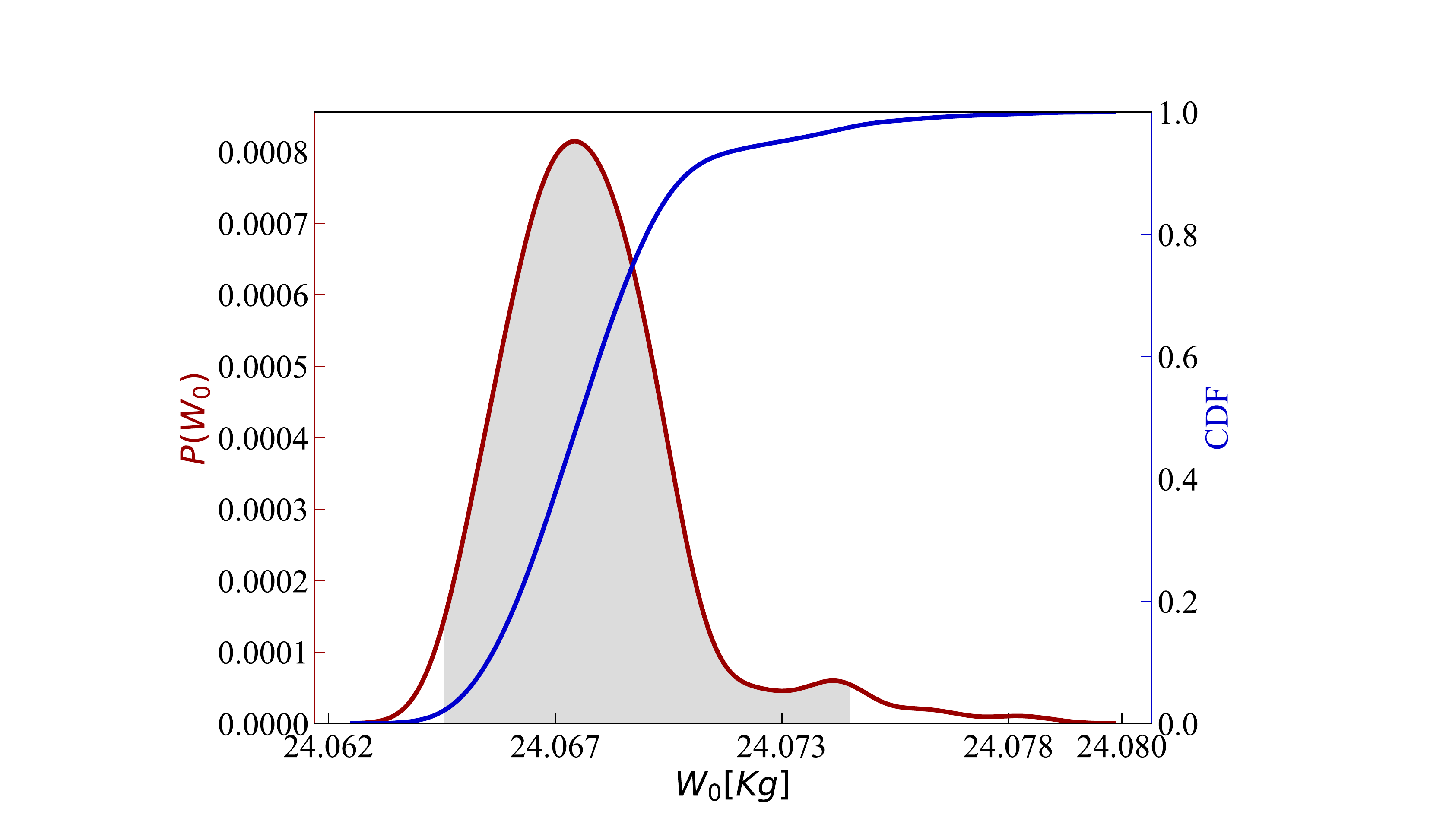}
\includegraphics[scale=0.275,trim=11cm -1cm 5cm 1cm, angle =0 ]{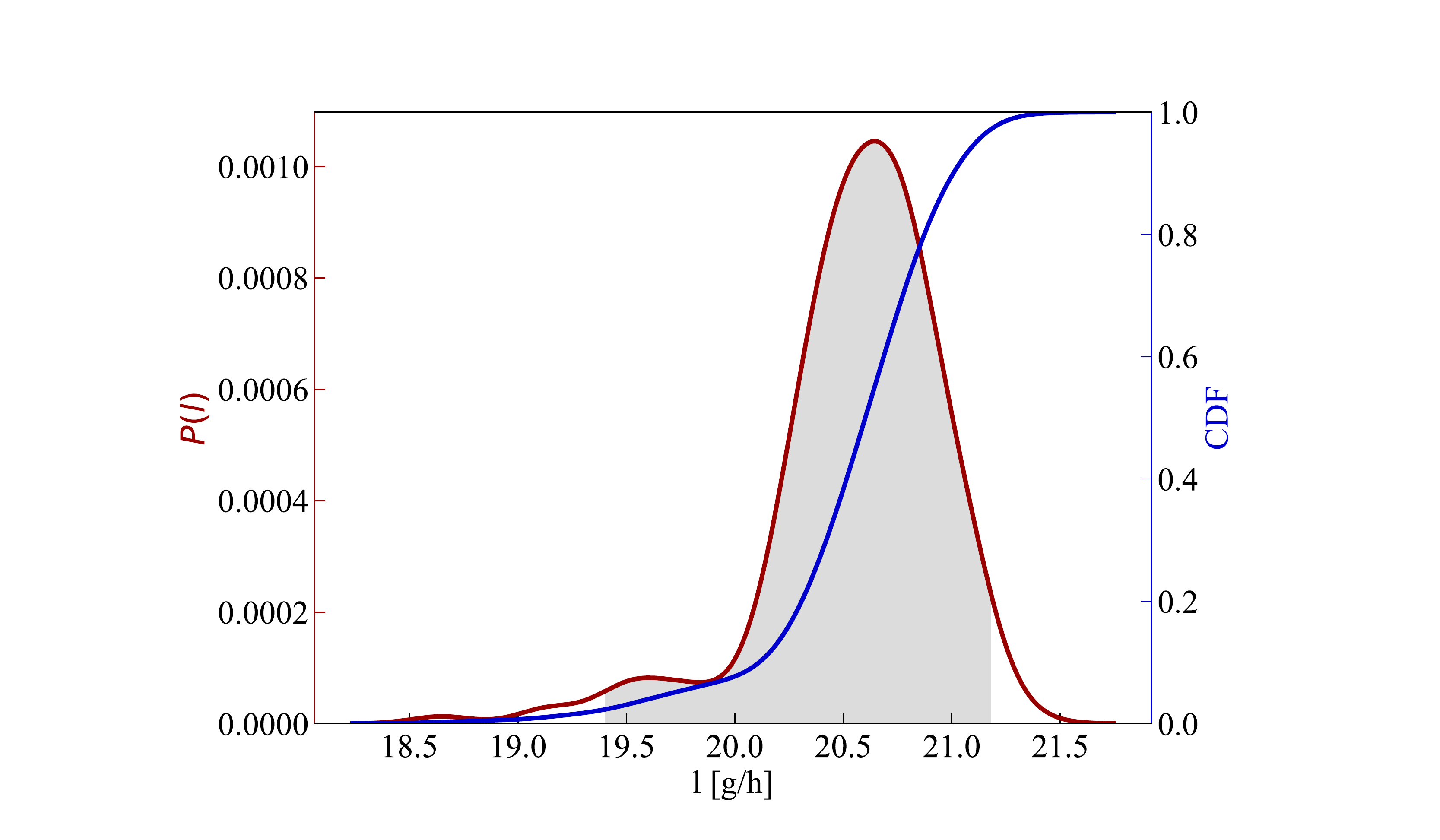}
\includegraphics[scale=0.275,trim=4cm -1cm 8cm 1cm, angle =0 ]{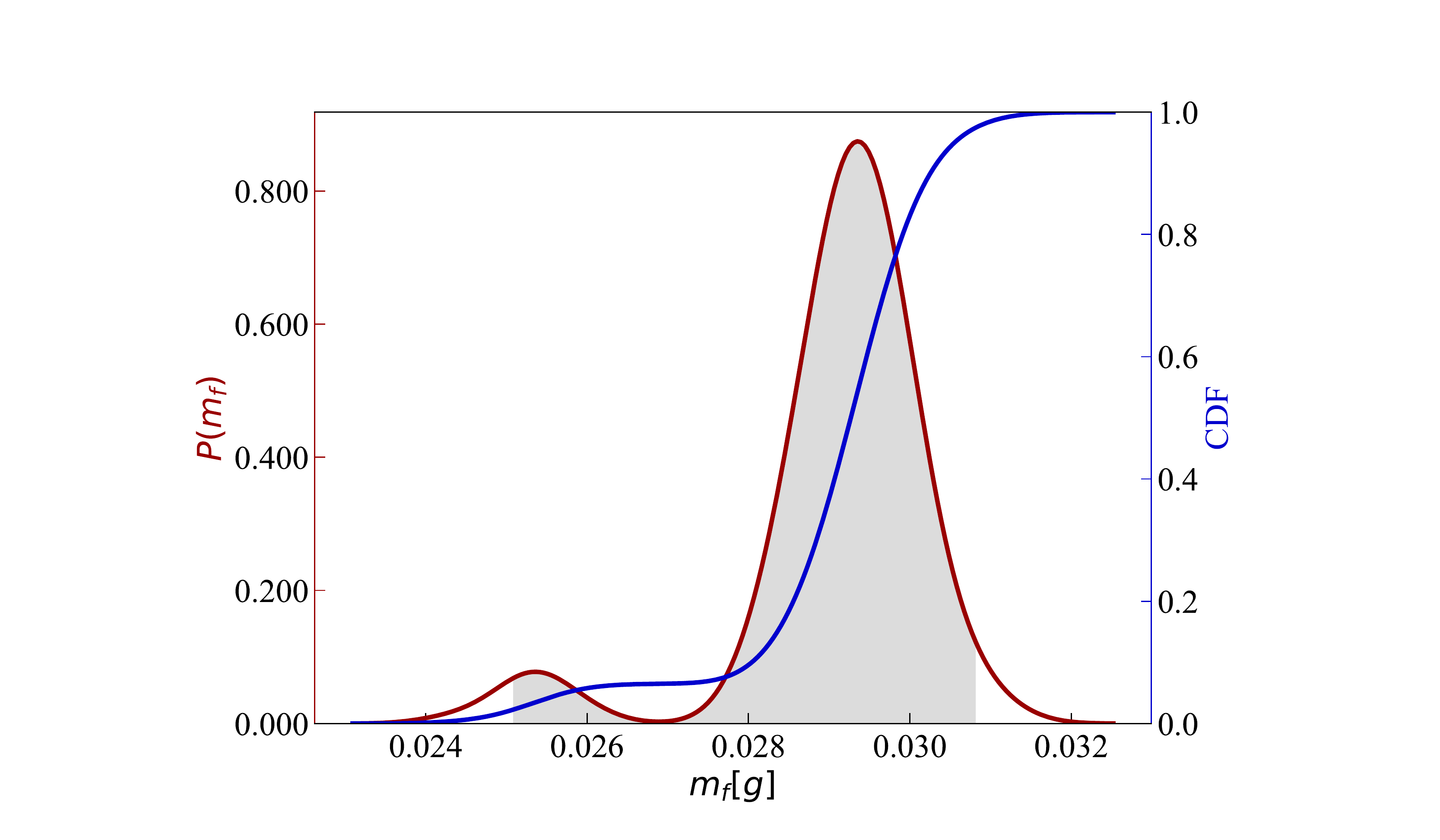}
\includegraphics[scale=0.275,trim=11cm -1cm 5cm 1cm, angle =0 ]{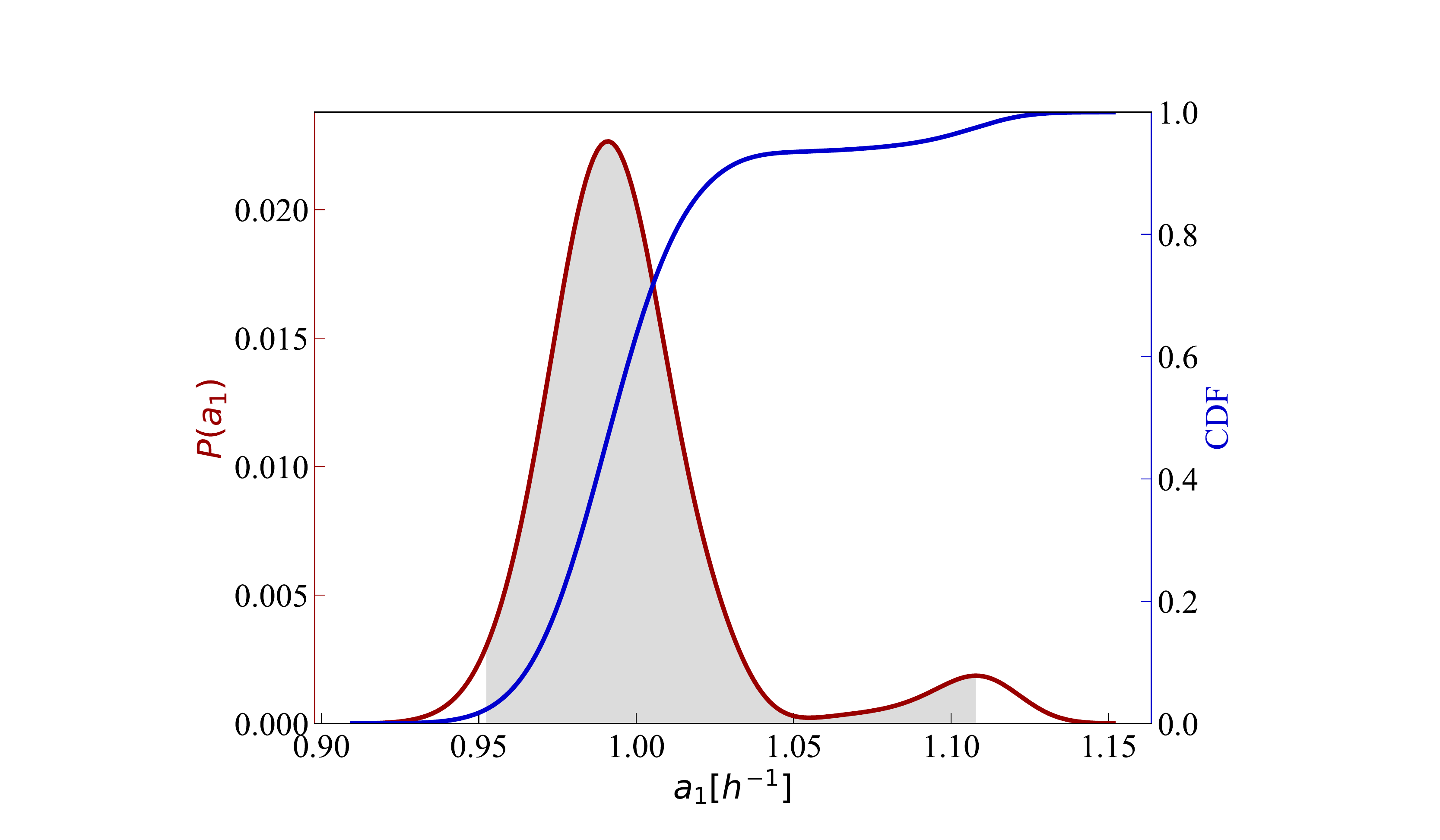}
\includegraphics[scale=0.275,trim=4cm -1cm 8cm 1cm, angle =0 ]{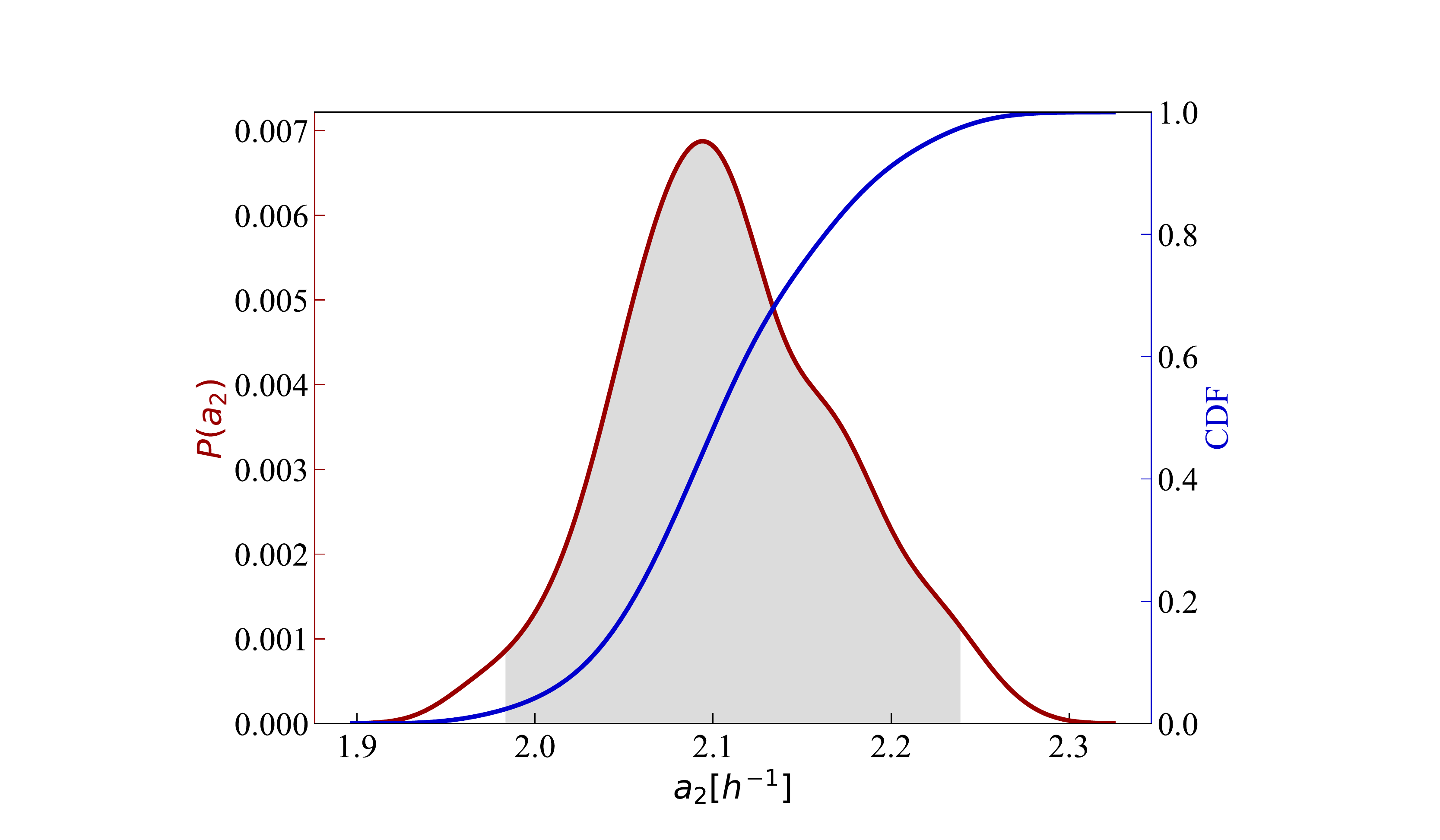}
\includegraphics[scale=0.275,trim=11cm -1cm 5cm 1cm, angle =0 ]{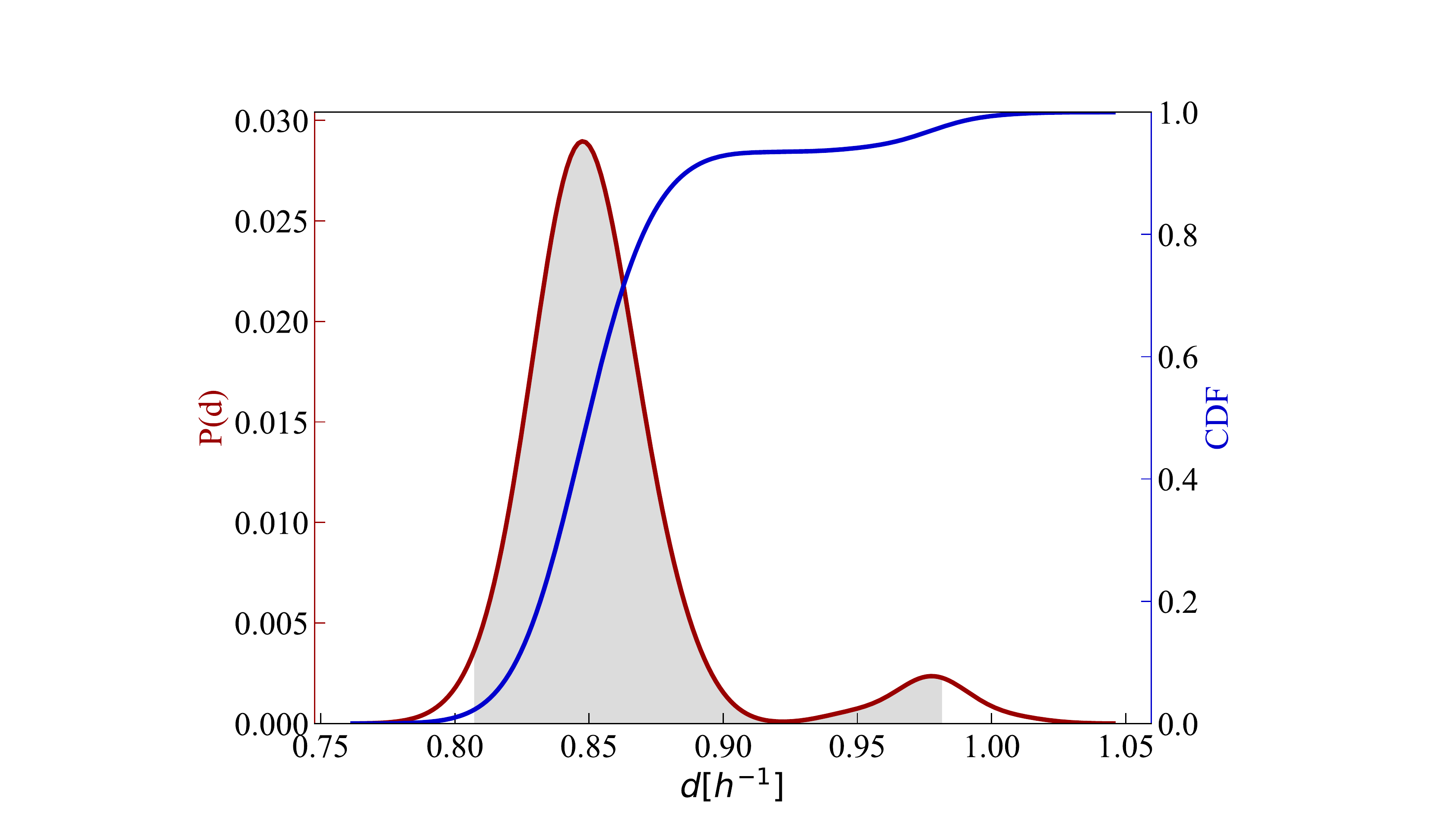}
\caption{ 
PDF for the fitting parameters: $N_{max}$[bees], $W_{0}$[kg], $m$[g], $a_1$[$h^{-1}$], $a_{2}$[$h^{-1}$], $d$[$h^{-1}$], $l$[g/h] with $t_{0}=8.4$h and $t_{1}=17.93$h after bootstrapping process of `Hive 6' during the 2018-4-7.} 
\label{fig.bootstrap}
\end{figure*}
\clearpage

\section*{Author Contribution}
Conceptualization: KAC, TC, MM, TL, EGA, Experiments: TC, Data Analysis: KAC, TC, Mathematical Model: KAC, TC, EGA, Coding: KAC, TC,  Writing – Original Draft Preparation: KAC, TC, EGA, and Writing – Review \& Editing: KAC, TC, MM, TL, EGA

\section*{Acknowledgement}
We acknowledge the Australian Research Council grant DP190101994. T.C. acknowledges funding from the Lord Mayors Charitable Foundation and Eldon and Anne Foote Trust. K.A.C. thanks The Sydney Informatics Hub at The University of Sydney for providing access to HPC-Artemis for data processing. We thank David Riley for helpful comments.

\bibliography{references}

\end{document}

%% file: Methods.tex
\section{Materials and Methods}
\subsection{Computational methods}\label{App.Computational_methods}
Here we describe the computational methods used to estimate ranges of the parameters $\theta$ of our model presented in Table \ref{tab.parameters} from the within day time series of weights $W'(t)$. In Sec.(\ref{sec.inference}) of this manuscript this issue has been formulated as the problem of finding the parameters $\theta$ that minimize the sum of squares~$L$ in Eq.~(\ref{Eq.Lsq}). The ten parameters proposed in this model were reduced to nine parameters after considering the weight of a bee $w$ as $0.113$ g/bee.

The minimisation was performed in a nine-dimensional space through a least-square method composed of three different approximation levels. First, the clock times when bees started departing $t_0$, and when they stopped leaving $t_1$. Next, a robust estimation of the characteristic features from the time series of the weight fluctuations was performed. Finally, the remaining seven parameters were obtained from the connection with the robust parameters. \par 
For the estimation of clock times for $t_0$ and $t_1$ we created a squared-grid for the times $t_{0}\in[$6:30am,9:00am$]$ and $t_{1}\in[$2:30pm,6:30pm$]$ spaced every $5$ minutes. These times combinations provides 1440 different scaling options. Then, we applied to each ordered pair, ($t_0,t_1$), the Levenberg\textendash Marquardt algorithm for square minimization. We used the python library SciPy (through ``scipy.optimize.curve\_fit" Pythons' function) with the inputs described in Tab.~\ref{tab.curve_fit} to apply the algorithm.

\begin{table}[!h]
\small
  \begin{tabular}{ll}
    \hline
\textbf{Parameter} & \textbf{Description}\\
$y_{data}:$& the measured weight $W'$ per minute over a full day\\
$x_{data}:$& the time per minute scaled for the corresponding measured weight $W'$\\
$f:$& the model function, introduced in Eq.~(\ref{eq.W})\\
$P_{0}:$& the initial guess for the $\theta$ parameters that are not bee-facts\\
& $W_{0}=15000 \, g, \,\,N_{max}=15000 \,\, \mathrm{bees}, \,\,l=17 g/h$\\
& $m=0.01 \,g/bee, a_{1}=0.21 \, h^{-1}$, $a_{2}=0.41 h^{-1}$, $\,\,d=1 h^{-1}$   \\
$Bounds$:& lower and upper bounds on parameters\\
& $W_{0}=[5000\quad max(W'))\, g, \:\:\: N_{max}=(0 \quad 80000] \,\, \mathrm{bees},  \qquad l=(0 \quad max(W')) g/s,$\\
& $\quad \quad \quad m=(0\quad 0.78w] \,g,\:\:\:\:\:\: \qquad  \qquad a_{1}=[0.10\quad 4.80] \, h^{-1}$,\, \\
& $\qquad \quad  a_{2}=[0.10\quad 4.80] h^{-1}$,  $d=[0.81 \quad 3.01] h^{-1}$  \\
$maxfev:$& the maximum number of evaluations allowed,\\
&for this function, 5000 iterations were used\\
$t_0,t_1:$& the clock time corresponding to $t_0$ and $t_1$, were obtained previously\\
\hline
  \end{tabular}
\caption{The input required for the ``scipy.optimize.curve\_fit" Pythons' function.}
\label{tab.curve_fit}
\end{table}
We choose the combination of $t_0$ and $t_1$ that minimizes the error as the inferred scale time parameters. Then, after re-scaling our data set on the time axis, we proceeded with the robust estimation of $A$, $B$, $t_c$, and $\alpha$ from the daily-hive data. We infered the remaining seven model parameters from a system of six equations, which consist of; (i) four equations presented as the ``effective description of the model". (ii) One equation from the conservation of the number of departed and returned bees. (iii) The last equation is the rate of evaporation $\ell$, which can be calculated directly from the time series after the re-scaling process when  $t<0$. Then, we performed a simultaneous calculation with six equations and seven variables; consequently, the system is undetermined with no unique solution due to an extra degree of freedom. We proposed that our dependent system of equations must be solved for any possible $N_{max}$ value. Note that the total weight of the hive at $W(t=0)$ can be decomposed in two terms only: the weight given by the total number of forager bees $N_{max}$, and the initial constant weight $W_{0}$. Consequently, our solution will depend on picking a value for $N_{max}$ and finding the corresponding $W_{0}$ value. Thus, an infinite number of solutions will be obtained and then reduced to a confidence interval as can be observed in Figure \ref{fig.6} in the Supporting information. 
Alternatively, by pre-defining one value, such as departure rate $d$, we can reduce the variables of our equation's system and obtain a unique solution for the parameters $\hat{\theta}$. The solution has extended to 10 independent and different hives during the day 2019-4-7 and is presented in Sec \ref{sec.results}.

\subsection{Pre-processing of data} \label{App.Data}
Before any computation, we removed outliers that do not reflect actual change in hive weight and that would otherwise affect our analysis. These outliers appeared due to equipment failure, animal or human interactions with the hives, or rainfall. In practice, we removed all measurement points that corresponded to a difference of 20g or more within two consecutive minutes, a conservative threshold which ensure that strong oscillations were excluded. From the ten different hives analysed, the hive that presented the highest number of outliers was hive `10', with  89 outliers, so at most only 6.1 \% of the data were excluded from this hive \dBlue{(Figure \ref{fig.outliers})}.\par

\subsection{Bootstrapping}\label{App.boostrapping}
Bootstrapping is used for assessing the effects of data uncertainty. We applied a bootstrapping method to estimate the uncertainty around the estimated parameters $\hat{\theta}$. For each daily data of 10 different hives, $200$ re-samplings were generated, and the minimisation procedure described above was applied to them (for the fixed value of $t_0,t_1$ obtained in the full dataset). Suppose the inference fails or the number of iterations exceeds $5,000$ evaluation, and the optimal parameters are still not found. In that case, the re-sample is discarded (for the 10 analysed hives listed in Tab.~\ref{tab.inferred} only 5\% of the 200 re-samplings were discarded). The inferred parameters of the accepted re-samples are used to estimate the range of plausible parameters. The uncertainty reported in the main manuscript corresponds to the standard deviation over the parameters estimated in the re-sampled data. For  $10$ hives reported in Tab.~\ref{tab.inferred} the uncertainty of the  7 inferred parameters corresponds to an area of 95.4 \% under the curve of the probability density function. Fig.~\ref{fig.bootstrap} in the Supporting information section displays an example of the probability density to calculate the range of values for the seven parameters of `Hive 6' on 2018-4-7.